\documentclass{aa}

\usepackage{graphicx,psfig,amssymb,float,natbib,rotating}
\bibpunct{(}{)}{;}{a}{}{,}

\begin{document}

\newcommand{\Po}{$P_{\rm orb}$}
\newcommand{\Pp}{$P_{\rm ph}$}
\newcommand{\e}{$e$}

\title{A Chandra observation of the old open cluster M\,67
\thanks{Tables~\ref{srclist} and \ref{fc} are only available in
electronic form via http://www.edpsciences.org} }

\titlerunning{A Chandra observation of M\,67}

\author{Maureen van den Berg\inst{1}\thanks{Present address:
        Harvard-Smithsonian Center for Astrophysics, 60 Garden Street,
        Cambridge, MA 02138, USA, maureen@head-cfa.harvard.edu} \and Gianpiero Tagliaferri\inst{1} \and
        Tomaso Belloni\inst{1} \and Frank Verbunt\inst{2}}
\authorrunning{Maureen van den Berg et al.}

\offprints{Maureen van den Berg}

\institute{ INAF/Osservatorio Astronomico di Brera, Via E. Bianchi 46,
              23807 Merate (LC), Italy;\\ vdberg@merate.mi.astro.it,
              tagliaferri@merate.mi.astro.it,
              belloni@merate.mi.astro.it 
              \and 
              Astronomical Institute,
              Utrecht University, P.O.Box 80\,000, 3508 TA Utrecht,
              The Netherlands; F.W.M.Verbunt@astro.uu.nl}

\date{Received date / Accepted date}   

\abstract{We present the results of a 47-ks Chandra-ACIS observation
of the old open cluster M\,67. We detected 25 proper-motion cluster
members (including ten new sources) and 12 sources (all new) that we
suspect to be members from their locations close to the main sequence
(1$<$$B$--$V$$<$1.7). Of the detected members, 23 are binaries. Among
the new sources that are members and probable members are four
spectroscopic binaries with $P_{\rm orb}$$<$12 d, two contact binaries
and two periodic photometric variables with $P_{\rm ph}$$<$8.4
d. Their X-rays are likely the result of coronal activity enhanced by
tidally locked rapid rotation.  The X-rays of the new source S\,997, a
blue straggler in a wide eccentric orbit, are puzzling.  Spectral fits
show that the X-rays of the brightest sources S\,1063 (a binary with a
sub-subgiant), S\,1082 (a triple blue straggler with a close binary)
and S\,1040 (a circular binary of a giant and a cool white dwarf), are
consistent with coronal emission. We detected a new bright source that
must have brightened at least about ten times since the time of the
ROSAT observations. It is not clear whether its faint blue optical
counterpart belongs to M\,67. We discuss the possibility that this
source is a low-mass X-ray binary in quiescence, which would be the
first of its kind in an open cluster. In addition to cluster members,
we detected about 100 background sources, many of which we identify
with faint objects in the ESO Imaging Survey.

}

\maketitle

\section{Introduction} \label{intro}

The X-rays of late-type stars are believed to result from a magnetic
dynamo driven by convection and rotation. As single stars spin down as
they age, their X-ray emission is expected to decrease
accordingly. Observations of young ($<$600 Myr) open clusters indeed
show a decline of the average X-ray luminosity with age
\citep{rand97}. Consequently, X-ray observations of {\em old} ($>$1
Gyr) open clusters are especially suited to look for special stars
among the cluster members, and above all to look for interacting
binaries.  The efficiency to detect binaries in old open clusters with
X-ray observations is illustrated by the results of ROSAT observations
of IC\,4651, NGC\,6940, NGC\,752, M\,67 and NGC\,188
\citep{belltagl,belltagl98,bellverb,bellea93,bellea}. In these
clusters, more than half of the X-ray sources with a cluster member as
optical counterpart, are identified with a binary of some sort. Mainly
they are active binaries in which tidal interaction forces the stars
into rotation that is synchronized to the orbit, thereby maintaining
rapid rotation even at high ages. Typical X-ray luminosities of
late-type active binaries in the ROSAT 0.1--2.4 keV band are
10$^{28}$--10$^{32}$ erg s$^{-1}$, where the lowest luminosities are
in general found for active main-sequence stars (BY\,Dra systems) and
the highest for active giants and subgiants \citep[RS\,CVn
binaries;][]{dempea,dempea3}. In this paper, we refer to both types as
RS\,CVn systems. The probability to detect the brightest systems in
young clusters -- with higher main-sequence turnoff masses -- is
smaller as more massive (sub)giants evolve more rapidly. Among the
counterparts in old open clusters are also binary blue stragglers and
a single, rapidly rotating FK\,Com giant (believed to be the product
of a merged binary).  Until now, ROSAT also detected one cataclysmic
variable (discovered in the optical) and one hot white dwarf, but due
to their optical faintness and the magnitude limits of optical
studies, such sources are much more difficult to identify.

Thus, in principle, observing old open clusters in X-rays is a way to
study the properties of their binary populations, characterized by the
clusters' properties like age and richness. Reversely, the number and
properties of the binaries in a cluster are important for the
dynamical evolution of a cluster \citep{hutea}.  X-ray observations
can be useful to point out close-binary candidates among stars that
are difficult to include in systematic radial-velocity searches for
binaries, for example among stars on the faint part of the main
sequence.

In practice, the number of old open clusters that is not too distant
or too reddened to be observed in X-rays, and for which the necessary
information exists to complement the X-ray data (like proper-motion,
radial-velocity and photometric data), is rather limited. One of the
few suitable clusters for an X-ray study is M\,67, a 4 Gyr-old cluster
that has been investigated in detail in the optical. M\,67 is a rich
cluster, is relatively nearby ($\sim$820 pc) and nearly unreddened
($E(B-V)=0.04(2)$), with a metallicity close to solar
\citep{saraea}. Proper-motion studies to establish cluster membership
have been done by e.g.~\citet{san}, \citet{giraea} and \citet{zhaoea};
in this paper we use membership information from \citet{giraea} that
reaches down to $V\approx16$.  ROSAT-PSPC observations of M\,67 are
presented by \citet{bellea93,bellea}.  The faintest source detected in
the combined ROSAT images has an X-ray flux of about 9~10$^{-15}$ erg
cm$^{-2}$ s$^{-1}$ (0.1-2.4 keV), which corresponds to a luminosity of
7~10$^{29}$ erg s$^{-1}$ at a distance of 820 pc. Here, we present the
results of a recent Chandra observation of M\,67 with a limiting flux
about 40 times lower.  With the increased sensitivity of this
observation compared to those of the ROSAT observations, we can detect
fainter sources, but single cluster stars remain out of
reach. Moreover, Chandra's subarcsecond positional accuracy avoids
confusion when looking for optical counterparts in the crowded field
of the cluster. We describe the data reduction and source detection in
Sect.~\ref{data}, the cross\-identification with optical and ROSAT
sources in Sect.~\ref{iden}, and the spectral and variability analyses
in Sect.~\ref{ana}. Sect.~\ref{disc} discusses individual systems in
more detail.

\section{Observation and data reduction} \label{data}

\subsection{Observation}
\begin{figure}
\centering
\resizebox{\hsize}{!}{\includegraphics{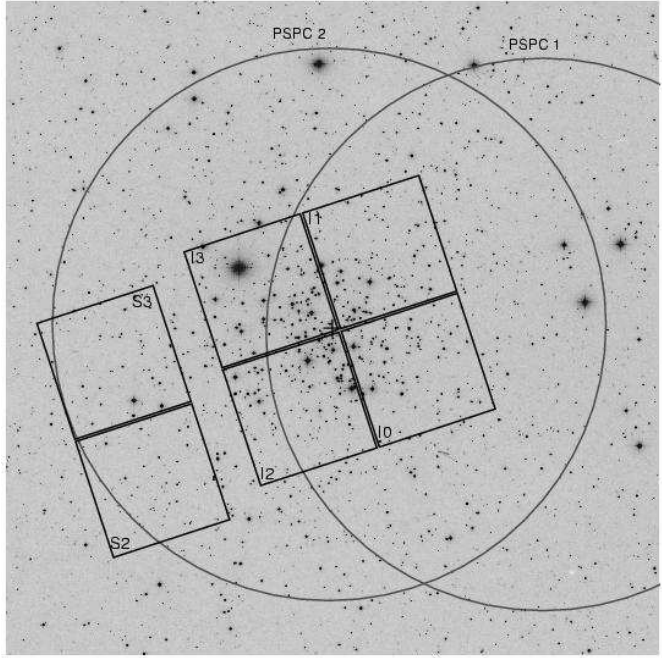}}    
\caption{A $45\arcmin \times 45\arcmin$ Digitized Sky Survey image of
M\,67 that shows the configuration of the six ACIS detectors exposed
in the observation. CCD numbers are indicated in the corners. The
ACIS-aimpoint was at $\alpha=$~8:51:23.1, $\delta=$~11:49:32.7
(\,J2000), marked with a cross on I\,3. Each CCD has a field of view
of $8\farcm4 \times 8\farcm4$. The positions of the inner ring of the
PSPC window support structure in the first and second ROSAT
observation are also indicated. North is up, east to the left.  }
\label{map}
\end{figure}

The central field of M\,67 was observed from May 31, 11:19 to June 1,
00:47 2001 UTC for 46.8 ks in timed exposure mode using faint
telemetry format (ObsId 1873) with the Chandra Advanced CCD Imaging
Spectrometer (ACIS, see {\em The Chandra Proposers' Observatory Guide}
for the instrument description).  The integration time of individual
frames was 3.2 s.  Each ACIS detector has a field of view of $8\farcm4
\times 8\farcm4$, while one pixel projects to $\sim$$0\farcs492 \times
0\farcs492$ on the sky.  The telescope aimpoint was at
$\alpha=$~8:51:23.1, $\delta=$~11:49:32.7 (J\,2000) on CCD I\,3, such
that the four ACIS-I detectors covered the cluster centre; two
additional ACIS-S CCDs (S\,2 and S\,3) imaged part of the eastern
regions of the cluster. See Fig.~\ref{map} for the configuration of
the detectors on the sky. The field covered by ACIS is fully included
in both ROSAT-PSPC images of the cluster.

\subsection{Data reduction}
Data reduction was done with the routines of the CIAO\,2.2 software
package following the threads on the CIAO webpages\footnote{see {\tt
http://asc.harvard.edu/ciao/}}. First, the level-1 events file was
processed with the task {\em acis\_process\_events} to make the
following corrections: application of a new calibration file ({\tt
geom.par}, released in version 2.9 of the Calibration Database);
removal of pixel randomization to improve the point spread function
(randomly spreading event-positions within one pixel is part of
standard data processing and is done to remove artificial patterns
caused by the satellite dither; in the image of M\,67, the pattern is
blurred automatically because the exposure time spans many
dither-cycles) and application of Pulse Height Amplitude randomization
(blurring of the energy values assigned to events to avoid that a
source's spectrum shows discrete peaks). We refer to the CIAO threads
for a more detailed description of these steps. A level-2 events file
was then generated in which only events with grades 0, 2, 3, 4 and 6,
and a status of 0 are preserved; ``flaring pixels'', contaminated by
cosmic-ray events, are filtered out in this step. An additional filter
is applied that only preserves data taken during good time
intervals. Finally, a correction is applied to the aspect solution
provided by the standard data processing.

Three images were created by splitting the observation in a soft
(0.3$-$2 keV), hard (2$-$7 keV) and total (0.3$-$7 keV) energy band.
The images were cleaned to remove time intervals with background
count\-rates that deviate more than $\pm3\,\sigma$ from the mean
value. As the CCDs have different characteristics, the cleaning was
done for each chip separately. This leads to a reduction of the
exposure time of $<2.5$\,\%. Next, an exposure map was computed for
each CCD in each energy band that represents the effective exposure of
each position on the detector by taking into account spatial
variations of the instrument's effective area and the dither motion of
the telescope.

\subsection{Source detection} \label{detect}
We used the {\em wavdetect} routine to detect sources in the unbinned
images. For ACIS-I, the detection was done on scales (wavelet radii)
ranging from 1 to 11.3 pixels, where scales are increased by
factors of $\sqrt{2}$. For ACIS-S, that lies further from the aimpoint
resulting in a larger point spread function, we used scales from 8 to
45.3 pixels. A small part of S\,2 is excluded from the detection as,
judging from the prominent square pattern, it appears to contain a bad
pixel. The significance threshold for detection was set to a 10$^{-6}$
probability per pixel for a spurious detection, corresponding to
$\sim$1 false detection per ACIS CCD of $1024 \times 1024$ pixels$^2$.
For some sources, {\em wavdetect} finds a very low number of
background-corrected counts (2$\pm$1); we do not reject these sources
as {\em wavdetect} is primarily designed for source detection and is
known to underestimate the source counts in some cases
\citep{freeea}. We computed vignetting-corrected countrates by
normalizing a source's exposure to the exposure near the aimpoint.

The resulting source lists of the different energy bands were
crosscorrelated by sky coordinates to create one master list, in which
sources are indicated with the prefix 'CX'. We adopt an error on the
position that is the quadratic sum of the 1-$\sigma$ {\em wavdetect}
errors, and a contribution that depends on the number of detected
counts (following information on the Chandra webpages, we add 0\farcs5
for sources with 20 counts or less, and 0\farcs2 for sources with 100
counts or less and more than 20 counts).  Sources are considered to be
a matching pair when their positions agree within these errors. A
first crosscorrelation results in a total of 193 sources. However,
after inspecting this list, we find that several sources are separated
by only 0\farcs05--3\farcs1. When increasing the {\em wavdetect}
errors to 2\,$\sigma$ and 3\,$\sigma$, and repeating the
crosscorrelation, we find 20 and 8 extra matches with respect to the
1\,$\sigma$ and 2\,$\sigma$ cases, respectively. There remain two
cases on ACIS-I in which sources lie close together and that we will
treat as matches: a hard-band source with 12 counts that lies 0\farcs9
from a total/soft-band pair (our final source CX\,10), and a soft-band
source with 4 counts that lies 1\farcs4 from a total-band source (our
final source CX\,114). We note that CX\,10 and CX\,114 could also be
blends of two different sources. There are five cases on ACIS-S, more
than $\sim$12\arcmin\,from the aimpoint, for which two sources of
different energy bands are too far apart to be a match, but whose
source regions (i.e.~the region with three times the size of the
standard deviation of the spatial distribution of source counts) for
the most part overlap. As their separations are small relative to the
sizes of the source regions, and as the uncertainty in Chandra
positions may be underestimated for large off-axis angles, we still
consider these detections as corresponding to the same X-ray source
(CX\,18, CX\,34, CX\,35, CX\,44, CX\,54). Our final master list
contains 158 sources. Table~\ref{srclist} lists the properties of the
sources, sorted on number of counts in the total band as derived by
{\em wavdetect}. In the following, we use positions derived from the
total-band image.

We now add some notes on individual sources. The source region of
CX\,54 includes the bad region on CCD S\,2 that was excluded from the
detection. We have not removed this source from our list, as a ROSAT
source lies in the source region of the Chandra source. We stress that
the properties of CX\,54 are likely not well represented by the
numbers given in Table~\ref{srclist}.  The source regions of CX\,32
(detected in all three bands) and CX\,45 (detected in soft and total
band) overlap; possibly {\em wavdetect} separated what looks like the
extended emission of the same source (the optical counterpart is an
active galaxy, see Sect.~\ref{opt}) in two.   Also the source regions
of CX\,71 (detected in total and soft band) and CX\,149 (detected in
hard band only) overlap; for this pair we only know that the optical
counterpart is faint and blue (see Sect.~\ref{opt}).

To get an idea about the spurious detections in our list, we repeated
the detection with the threshold set to a 10$^{-7}$ probability per
pixel for a spurious detection. The 26 sources that are not detected
in this run of {\em wavdetect} but that are found with the threshold
set to 10$^{-6}$, are marked with an asterisk ($^{*}$) in
Table~\ref{srclist}.

\begin{table*}
\caption{X-ray sources in the field of M\,67 detected by Chandra.  For
each source we list the Chandra number, the distance from the
aimpoint, the celestial position (J\,2000) derived from the image in
the total energy band, the number of background-corrected counts in
the total energy band derived by {\em wavdetect}, the net countrate
(i.e.\,corrected for background counts and for vignetting) in the
total band, the energy band $B$ in which the detection has the highest
significance (S=0.3$-$2 keV, H=2$-$7 keV, T=0.3$-$7 keV), and the
hardness ratio $HR$ (if $\sigma_{HR} < 0.4$). If appropriate, we also
give the ROSAT counterpart 'X' \protect{\citep{bellea}}, the optical
counterpart 'opt', and the distance $d_{\rm XO}$ between the Chandra
and the optical source. The name of the optical counterpart is
selected from (in order of preference): \protect{\citet{san}} (S), the
EIS list of \protect{\citet{momaea}} (E), and \protect{\citet{fanea}}
(F). Sources that are identified by eye with objects listed in neither
catalogue are labelled with 'faint' (see Table~\ref{fc}).  {\em Only
the first lines are shown here to demonstrate the table format. The
full table is available only in the electronic version of the paper.}}
\begin{tabular}{rrllrcrrlll}
\hline \hline CX  & $d_{\rm aim}$ &
\multicolumn{1}{c}{$^{**}\alpha_{\rm J2000}$}  &
\multicolumn{1}{c}{$^{**}\delta_{\rm J2000}$} & counts & ctr & $B$ &
\multicolumn{1}{c}{$HR$} & X   & opt  & $d_{\rm XO}$  \\ &
\multicolumn{1}{c}{(\arcmin)}  &  \multicolumn{1}{c}{($^h$,$^m$,$^s$)}
&  \multicolumn{1}{c}{($^{\circ}$,\arcmin,\arcsec)}   &    &
(c\,ks$^{-1}$) &  &  &      &      & (\arcsec) \\ \hline 1 &  3.19   &
8 51 13.373(2)      & 11 51 40.23(2)  & 892(30)  & 20.5 &  T &
0.45$\pm$0.03   &  8   & S\,1063    & 0.27 \\ 2 &  3.78   &   8 51
38.456(1)    &   11 49 06.28(3)  & 872(30)  & 21.5 &  T &
0.36$\pm$0.03   &      & E\,510     & 0.13 \\ 3 &  3.93   &   8 51
20.798(2)    &   11 53 26.40(3)  & 619(25)  & 14.3 &  T &
0.37$\pm$0.04   &  4   & S\,1082    & 0.23 \\ 4 &  7.23   &   8 50
57.104(8)    &   11 46 07.57(8)  & 513(24)  & 14.5 &  T &
0.65$\pm$0.04   & 15$^{c}$ & faint  & 0.1 \\
\hline
\end{tabular}

$^*$ Source is detected with the significance threshold set to
$10^{-6}$ but not with the threshold set to $10^{-7}$\\ $^{**}$ All
positions are corrected for a systematic offset with respect to the
EIS optical positions which is $\Delta \alpha = \alpha_{\rm X} -
\alpha_{\rm opt} = -$0\farcs23 and $\Delta \delta = \delta_{\rm X} -
\delta_{\rm opt} = -$0\farcs20. The errors on the positions that we
give here are the 1-$\sigma$ {\em wavdetect} errors including the
corrections for small counts numbers. They are often much smaller than
the uncertainty in the absolute positions, i.e.\, than the uncertainty
in the offset values ($\sim$0\farcs3).\\
\label{srclist}
\end{table*}

\section{Crossidentification with optical and ROSAT sources} \label{iden}

\subsection{Optical identification} \label{opt}

\begin{figure}
\resizebox{\hsize}{!}{\includegraphics[clip]{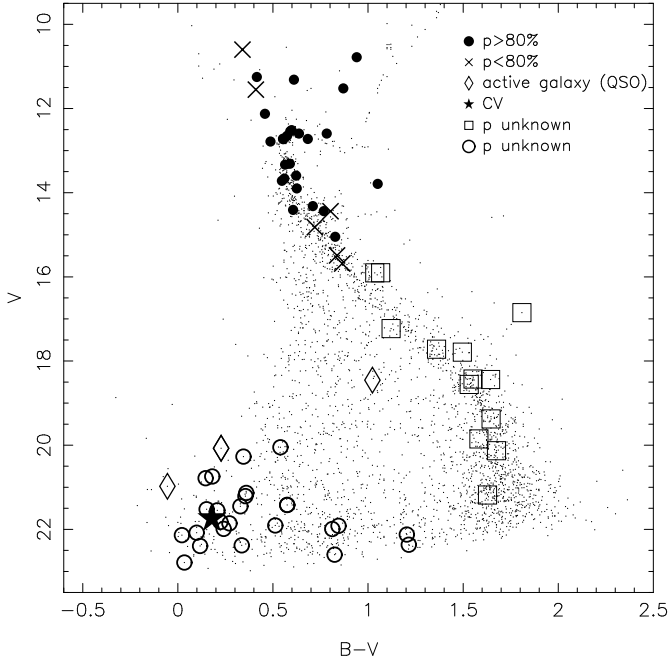}}
\caption{Colour-magnitude diagram of M\,67 that shows the optical
counterparts of the Chandra sources listed in Tables~\ref{optid} and
\ref{optid-2}. The meaning of the symbols is included in the figure,
where $p$ is the proper-motion probability for cluster membership, and
CV is the cataclysmic variable EU\,Cnc. Sources without proper motion
information are divided into sources near (squares) and to the blue of
(circles) the main sequence.}
\label{cmd}   
\end{figure}

To look for optical counterparts of the Chandra sources, we
crosscorrelated our source list against several optical
catalogues. Our main reference for objects with $11 \lesssim V
\lesssim 23$ is the catalogue from the ESO Imaging Survey (EIS, Momany
et al. 2001). \nocite{momaea} Before crosscorrelating the X-ray and
optical lists, we had to correct for any systematic offset between the
X-ray and the optical positions, which we have done iteratively.  We
started by overplotting the X-ray sources on the EIS image and thus
identified by eye optical counterparts for the brightest sources.
From the positions of twelve stars that were already considered to be
X-ray sources from ROSAT and optical follow-up observations, we
computed median offsets for right ascension and declination. After
correcting the X-ray positions for these offsets, we correlated the
X-ray and optical lists. When an X-ray and optical position match
within the errors, the optical source is accepted as a
counterpart. For the X-ray sources, we adopted the same errors (with
the 1-$\sigma$ {\em wavdetect} contributions) used to correlate the
source lists of the three energy bands. We used the maximum internal
error on the EIS position ($\sim$1\arcsec, see Fig.~3 of
\citet{momaea}) as the error on the optical positions. With the new,
larger set of counterparts, we calculated a new offset between the
X-ray and optical positions and again correlated the lists after
applying the updated correction. This was repeated until no new
counterparts are found. In this way, we have found 56 EIS optical
counterparts. The positions in Table~\ref{srclist} are corrected for
the final offsets: $\alpha_{\rm X} - \alpha_{\rm opt}
=-$0\farcs23$\pm$0\farcs35 and $\delta_{\rm X} - \delta_{\rm opt} =
-$0\farcs20$\pm$0\farcs33. The precision of the X-ray positions is
limited by the uncertainty in these offsets.  For ACIS-S sources, we
possibly underestimated the positional errors, and therefore we
repeated the crosscorrelation using 2-$\sigma$ and 3-$\sigma$ {\em
wavdetect} errors.  Since in five cases we associated soft- and
total-band sources with a relatively large separation (see
Sect.~\ref{detect}), the crosscorrelation was done for each energy
band separately in order not to miss possible counterparts. We found
four (three) extra counterparts with 2-$\sigma$ (3-$\sigma$) errors.
To look for extra counterparts, e.g.~stars that are too bright to be
included in the EIS list, we repeated the above steps with the
catalogues of \citet{montea}, \citet{giraea}, and \citet{fanea}. We
thus found five more counterparts.  For each identification, the
distance between the X-ray source and optical counterpart is given in
Table~\ref{srclist}, in which the notes indicate whether the
counterpart was found using larger errors or the additional
catalogues. Tables~\ref{optid} and \ref{optid-2} summarize the
properties of the counterparts, while Fig.~\ref{cmd} shows their
positions in the M\,67 colour-magnitude diagram. We use the photometry
from EIS for stars with $V>14$ and from \citet{montea} for the
remaining stars, as the EIS magnitudes of bright stars can be affected
by saturation effects.

All identifications were checked by eye.  The errors on the X-ray
positions are so small that we either find zero or one possible
counterpart. A known QSO lies between CX\,32 and CX\,45, where their
source regions overlap (see Sect.~\ref{detect}); we tentatively
associate this QSO with both X-ray sources. Similarly, we tentatively
match the faint blue counterpart of CX\,71 with the nearby source
CX\,149 (see Sect.~\ref{detect}.)  We indicated eight more optical
sources as possible counterparts for X-ray sources on ACIS-I; they lie
outside the errorboxes, but inside the X-ray source regions and only
1\arcsec--4\arcsec\,from the X-ray sources. Possible counterparts are
indicated in {\em italics} in Tables~\ref{optid} and
\ref{optid-2}. Finally, we found fifteen possible counterparts by
examining the EIS $V$-image to look for faint optical sources that are
not included in either catalogue. Their estimated positions are given
in Table~\ref{fc}.

\begin{table}[!ht]
\caption{Properties of optical counterparts with proper-motion
probabilities for membership $p$$>$80\,\% (top) and $p<80$\,\%
(bottom). The first columns list the Chandra source (CX) and, if
appropriate, the corresponding ROSAT source (X). Optical sources are
identified with their number from \citet{san}; an 'S' indicates that
the present counterpart is selected from several ROSAT candidate
counterparts.  The next columns give $V$, $B-V$, the unabsorbed X-ray
luminosity $L_{\rm X}$ in the 0.3--7 keV band (in erg s$^{-1}$) ($p$
for probable non-members), the orbital period $P_{\rm orb}$ (in days;
'var' denotes a radial-velocity variable) and eccentricity
$e$. References for information on binarity (see notes), and remarks
are given in the last column (SG=sub-subgiant, BS=blue straggler,
YS=yellow straggler, WU=contact binary of W\,UMa type, bin=star lies
on binary sequence). Identifications that are uncertain because of
relatively large offsets between X-ray and optical positions are
printed in {\em italics}; see the discussion in Sect.~\ref{opt}.  The
conversion of counts to flux is discussed in Sect.~\ref{hard}.}
\label{optid}
\begin{tabular}{r@{\hspace{0.21cm}}r@{\hspace{0.21cm}}l@{\hspace{0.21cm}}l@{\hspace{0.21cm}}l@{\hspace{0.21cm}}l@{\hspace{0.21cm}}l@{\hspace{0.22cm}}l@{\hspace{0.22cm}}l}
\hline
\hline
CX & X & \multicolumn{1}{l}{S} &  $V$ & $B$--$V$ & $L_{\rm X}$/$p$ & $P_{\rm orb}$ & $e$ & rem. \\
\hline
  1 &  8 & 1063 & 13.79 & 1.05 & 1.3~10$^{31}$ & 18.40 & 0.21 & 1,\,SG \\ 
  3 &  4 & 1082 & 11.25 & 0.42 & 8.5~10$^{30}$ & 1.07$^{a}$ & & 2,\,BS \\ 
  5 & 11 & 1019 & 14.32 & 0.71 & 6.0~10$^{30}$ & 1.36 & 0 & 3 \\ 
  6 & 10 & 1040 & 11.52 & 0.87 & 3.3~10$^{30}$ & 42.83 & 0 & 4,\,YS \\ 
  9 & 13 & \mbox{~\,}999 & 12.60 & 0.78 & 3.7~10$^{30}$ & 10.06 & 0 & 4  \\ 
 10 &  7 & 1077 S & 12.60 & 0.64 & 3.9~10$^{30}$ & 1.36$^{a}$ & 0.10 & 3 \\ 
 16 & 40 & 1282 & 13.33 & 0.56 & 1.5~10$^{30}$ & 0.36 & & 5,\,WU \\ 
 19 & 45 & 1036 & 12.78 & 0.49 & 1.2~10$^{30}$ & 0.44 & & 8,\,WU \\ 
 23 &    & \mbox{~\,}757 & 13.59 & 0.62 & 8.5~10$^{29}$ & 0.36 & & 9,\,WU \\ 
 24 & 37 & 1072 & 11.32 & 0.61 & 7.6~10$^{29}$ & 1495 & 0.32 & 4,\,YS \\ 
 36 & 53 & 1234 S & 12.65 & 0.57 & 6.8~10$^{29}$ & 4.36$^{a}$ & & 4 \\ 
 47 & 52 & 1237 & 10.78 & 0.94 & 5.5~10$^{29}$ & 698 & 0.11 & 4,\,YS \\    
 48 & 38 & 1070 & 13.90 & 0.63 & 3.1~10$^{29}$ & 2.66 & 0 & 6  \\ 
 49 & 50 & 1242 & 12.72 & 0.68 & 2.9~10$^{29}$ & 31.78 & 0.66 & 4  \\ 
 67 &    & {\em 1203} & {\em 14.44} & {\em 0.77} &  {\em 3.6~10$^{29}$} & & & bin. \\ 
 72 &    & \mbox{~\,}773 & 13.31 & 0.59 & 1.6~10$^{29}$ & 5.7 & & 3 \\ 
 78 &    & 1009 & 13.67 & 0.56 & 3.0~10$^{29}$ & 5.95 & 0 & 6 \\ 
 81 &    & \mbox{~\,}996& 15.05 & 0.83 & 1.2~10$^{29}$ & 6.7   & & 3 \\ 
 88 & 41 & 1045 & 12.54 & 0.59 & 9.2~10$^{28}$ & 7.65 & 0 & 4 \\ 
 94 &    & 1281 & 13.72 & 0.55 & 1.6~10$^{29}$ &  & & \\  
 95 &    & \mbox{~\,}997 & 12.13 & 0.46 & 1.2~10$^{29}$ & 4913 & 0.34 & 7,\,BS \\ 
104 & 41 & 1050 & 14.41 & 0.61 & 1.1~10$^{29}$ & var & & 3 \\  
111 & 46 & 1024 S & 12.72 & 0.55 & 8.0~10$^{28}$ & 7.16 & 0 & 4 \\  
155 &    & {\em 1272} & {\em 12.51} & {\em 0.60} & {\em 4.2~10$^{28}$} & {\em 11.02} & {\em 0.26} & {\em 4} \\  
157 &    & \mbox{~\,}986 & 12.73 & 0.55 & 2.2~10$^{28}$ & 10.34 & 0 & 4 \\ 
\hline
  7 & 47 & 1601 & 14.44 & 0.80 &        0.48 & &  &   \\
 15 & 17 & \mbox{~\,}972 & 15.49 & 0.84  & 0.27 & 1.17 & 0 & 3  \\
 17 & 42 & 1042 & 15.68 & 0.86 & 0. & &   &   \\
 68 & 44 & 2214 S & 14.82 & 0.72 &       0. &  &  &  \\
109 &    & 1466  & 10.60  & 0.34 &       0.21 & & &  \\  
118 &    & 1013  & 11.55  & 0.41 &       0.  &  & &  \\  
\hline
\end{tabular}
$^{a}P_{\rm orb}$ refers to the inner orbit of a (possible) triple system.\\
1=Mathieu~et~al.\,(2003),~2=Goranskij~et~al.\,(1992),
3=preliminary~solution~(D.\,Latham,\,priv.\,comm.), 4=Mathieu et
al. (1990), 5=Kurochkin (1960), 6=Latham et al.~(1992), 7= Latham \&
Milone (1996), 8=Gilliland et al.~(1991), 9=Stassun et al.~(2002)
\end{table}
\nocite{mathea} \nocite{goraea} \nocite{kuro} \nocite{mathlathea} \nocite{lathmathea} \nocite{gillea} \nocite{lathmilo} \nocite{stasea2002}

\nocite{mathlathea,stasea2002}

\begin{table}
\caption{Properties of optical counterparts without proper-motion
information. The top part refers to sources near the main sequence,
the lower part to the remaining sources. The first two columns list
the Chandra source (CX) and, if appropriate, the corresponding ROSAT
source (X). Optical sources are indicated with their EIS numbers; we
indicate whether the identification is new (N), or is a correction of
a ROSAT identification (C).  References (see Table~\ref{optid}), and
remarks are given in the last column (CV=cataclysmic variable,
WU=contact binary of W\,UMa type, bin=star on binary sequence, $P_{\rm
ph}$ = photometric period; 'F' and 'LB' refer to sources from
\protect{\citet{fanea}} and
\protect{\citet{luyt,luyt2}}). Identifications that are uncertain
because of relatively large offsets between X-ray and optical
positions are printed in {\em italics}; see the discussion in
Sect.~\ref{opt}. Uncertain identifications with ROSAT sources are
marked with a '?'; see Sect.~\ref{rosatid}.}
\label{optid-2}
\begin{tabular}{r@{\hspace{0.2cm}}c@{\hspace{0.22cm}}l@{\hspace{0.22cm}}l@{\hspace{0.22cm}}l@{\hspace{0.18cm}}l}
\hline
\hline
CX    & X & \multicolumn{1}{l}{E}    & $V$    & $B$--$V$ & remarks \\ 
\hline 
 54   & 56? & \mbox{~\,}207 & 16.85 & 1.81 & near bad CCD-region\\
 58   &     &  \mbox{~\,}763 & 15.90 & 1.04 & 9, $P_{\rm ph}$=3.58 d; bin. \\
 61   &     &  \mbox{~\,}742 & 15.90 & 1.07 & 8, $P_{\rm ph}$=0.27 d; WU \\
 62   &     & {\em \mbox{~\,}683} & {\em  18.44} & {\em  1.64} & \\
 73   &     & {\em 2650} & {\em 21.18} & {\em 1.63} &  \\
 76   &     & 1720 & 17.72 & 1.36 & 9, phot. variable, bin.\\ 
 77   &     &  \mbox{~\,}394 & 19.38 & 1.65 & bin.\\ 
 80   &     & 2228 & 17.23 & 1.12 & \\
 82   &     & 1208 & 18.56 & 1.53 & \\  
117   &     &  \mbox{~\,}542 & 20.13 & 1.68 & \\  
129   &     & {\em \mbox{~\,}556} & {\em 17.79} & {\em 1.50} & \\
141   &     &  \mbox{~~~}--    & 19.85 & 1.58 & 8, $P_{\rm ph}$=8.4 d; F\,3925 \\
153   &     & 1178 & 18.43 & 1.55 & bin. \\
\hline
  2  &     &  \mbox{~\,}510 & 20.88 & 0.15 & \\
 12  &     & 2024 & 22.50 & 0.12 & \\
 13  & 14? & 2153 C & 21.21 & 0.36 & LB\,6322\\
 14  & 34  & 2823 C & 21.99 & 0.81 & \\
 20  & 43  & 2757 C & 20.98 &--0.055 & LB\,6378, QSO \\
 21  & 49  & 2183 C & 21.86 & 0.27 & \\
 22  &     & 2756 & 22.12 & 1.20 & \\
 26  & 55  & 2018 N & 21.52 & 0.15 & \\
 28  &     & 2297 & 22.37 & 1.21 & \\
 30  &     & \mbox{~\,}210  & 20.05 & 0.54 & \\
 31  & 48  & 2190 N & 22.38 & 0.34 & \\
 \multicolumn{1}{l}{32/45}  & 39  & {\em 1268} & {\em 20.07} & {\em 0.23} & LB\,3597,\,QSO \\
 33  &     &  \mbox{~\,}600 & 20.74 & 0.18 & LB\,6374  \\
 34  &     & 2722 & 22.08 & 0.099 & \\
 35  &     & 1929 & 21.92 & 0.85 & \\
 37  &     & 3674 & 22.83 & --  & \\ 
 50  &     & 1939 & 21.14 & 0.36 &  \\
 53  & 24  & {\em  \mbox{~\,}362 C} & {\em 20.27} & {\em 0.35} & LB\,6316 \\
 55  &     & 2684 & 21.55 & 0.21 & LB\,6364 \\ 
 56  &     & 2050 & 22.14 & 0.021 & \\
 57  & 16  & 2164 & 21.75 & 0.16 & 8, $P_{\rm ph}$=2.1 h; CV  \\
 64  &     & 2306 & 21.45 & 0.33 &  \\
 65  &     & 2307 & 22.60 & 0.82 &  \\
 70  &     & 2208 & 22.79 & 0.034 &  \\
 71  &     & 2807 & 21.42 & 0.58 & see also CX\,149 \\
 79  &     & 2310 & 21.82 & 0.22 &  \\ 
 86  &     & {\em 2135} & {\em 21.99} & {\em 0.24} & \\
 89  & 38  & 2270  & 21.91 & 0.51 &  \\  
100  &     & 3668 & 22.81 & --      &  \\  
127  &     & 3701 & 21.71 & --      &  \\  
\hline     
\end{tabular}
(continued on the right)
\end{table}

\setcounter{table}{2}
\begin{table}
\caption{continued}
\begin{tabular}{rcllll}
\hline
\hline
CX    & X & \multicolumn{1}{l}{E}    & $V$    & $B$--$V$ & remarks \\ 
\hline 

139  &     & \mbox{~~~}--     & 19.33$^b$ & --      & F\,4345 \\
149  &     & {\em 2807} & {\em 21.42} & {\em 0.58} & see also CX\,71\\
152  &     &\mbox{~~~}--     & {\em 18.45} & {\em 1.02} & F\,2776, LB\,3596; QSO \\
158  &     & 3534 & 20.09 & --      &  \\
\hline     
\end{tabular}
$^b$magnitude at 6075 \AA.
\end{table}

\subsection{Identification with ROSAT sources} \label{rosatid}

Of the 59 ROSAT sources detected by \citet{bellea}, 24 lie outside the
ACIS field of view.  In what follows, we indicate ROSAT sources with
an 'X'. Of the 35 sources in the ACIS field, 26 have one Chandra
counterpart in or just outside ($<$\,0\farcs3) their 90\% error
circles, while for five we find two counterparts: X\,38 (CX\,48,
CX\,89), X\,39 (CX\,32, CX45), X\,46 (CX\,111, CX\,112), X\,53
(CX\,36, CX\,120; both near the border of the error circle), and X\,41
(CX\,88, and CX\,104 just outside the error circle). Three possible
identifications are X\,56/CX\,54, X\,14/CX\,13 and X\,15/CX\,4 for
which the Chandra sources lie 2--4\arcsec\,outside the ROSAT error
circles.  These (tentative) crossidentifications are included in
Tables~\ref{srclist}--\ref{fc}.  The closest Chandra source to the
remaining ROSAT source X\,54 is CX\,56, at a distance of 25\farcs9 or
2.5 times the ROSAT error radius; we do not consider it a counterpart.

As Chandra's spatial resolution is many times better than ROSAT's, we
can check if the optical identifications suggested by \citet{bellea}
still hold.  In four cases we find a new counterpart (the deep EIS
list was only recently published), in four cases we can distinguish
between several suggested counterparts, and in five cases we find a
different counterpart (see Tables~\ref{optid}--\ref{fc}). We exclude
the identification of CX\,11/X\,51 with star 6441 from \citet{montea}
and of CX\,44/X\,58 with the probable non-member S\,1414; we find no
alternative counterparts. The number of corrected ROSAT
identifications is in agreement with the estimated number of chance
coincidences in \citet{bellea}.

\subsection{Spurious identifications}
We now estimate the probability that an X-ray source and an optical
EIS-source are crossidentified by chance. This probability depends on
the area (error box) around the X-ray source within which we search
for counterparts, and on the surface density of optical sources. Since
both the average errors on the X-ray positions ($\delta$) and the
stellar surface density vary with distance from the aimpoint at the
cluster centre, we make an estimate for ACIS-I ($\delta$=0\farcs44)
and ACIS-S ($\delta$=1\farcs1) separately. We use a box with sides
equal to twice the quadratic sum of $\delta$ and the error on the
optical positions (1\arcsec), to compute the search area around an
X-ray source. The number of EIS-sources inside ACIS-I and ACIS-S is
1264 and 472\footnote{Although the field where Chandra detected
sources is contained within the EIS image, the S\,2 and S\,3 CCDs are
not fully covered by the EIS image. Therefore, the number of 472 that
we use here, is twice the number of optical sources within S\,2, for
which only $\sim$0.5\arcmin$^2$ are outside the EIS field of view.},
respectively. Therefore, the probability that a randomly placed box
includes an EIS-source is (2 $\times$ 1\farcs1)$^2$$\times
(1264/(4\cdot(8.4\cdot60'')^2))=6.0 \cdot 10^{-3}$ for ACIS-I, and (2
$\times$ 1\farcs5)$^2$$\times (472/(2\cdot(8.4\cdot60'')^2))=8.4 \cdot
10^{-3}$ for ACIS-S.  We detected 140 X-ray sources on ACIS-I; thus we
compute that for 140 trials, the probability to identify an X-ray
source with 0, 1 or 2 EIS-sources by chance is 43, 36 and 15\,\%,
respectively.  The corresponding numbers for ACIS-S (18 sources) are
86, 13 and 0.9\,\%. In a similar way, we compute the probability for
spurious identifications with proper-motion cluster members, and with
proper-motion members that are known binaries. On ACIS-I, 230 stars
have a membership probability $p>80$\,\%, while there are 48 cluster
binaries (see \citet{lathmathea}, \citet{lathmilo},
\citet{vdbergea2001ad}, four contact binaries from \citet{gillea} and
\citet{stasea2002}, and seven binaries with unpublished orbits
(D.~Latham, priv.~comm.)). On ACIS-S, there are 14 stars with
$p>80$\,\%, while there are no known binaries. The results are
summarized in Table~\ref{spurious}. From these numbers we conclude
that probably all identifications with cluster binaries (see
Table~\ref{optid}) are correct.

We also estimate the probability for a chance identification of a
Chandra source with a ROSAT source. Using the average 90\,\% error
radius of the ROSAT sources inside ACIS (11\farcs6), we find that for
one trial this probability is $1.2~10^{-2}$ and $5.0~10^{-3}$ on
ACIS-I and ACIS-S, respectively. The probabilities for one or two
chance superpositions on ACIS-I are relatively large
(Table~\ref{spurious}), so some of our 34 identifications are likely
to be false.

\setcounter{table}{4}
\begin{table}[!t]
\caption{Probabilities for 0, 1 and 2 spurious crossidentifications of
Chandra sources with sources in the EIS catalogue, with stars that
have a probability for cluster membership $p>80$\,\%, with binaries
with $p>80$\,\% (no such binary inside ACIS-S), and with ROSAT
sources.}
\label{spurious}
\begin{tabular}{@{\hspace{0.1cm}}l@{\hspace{0.35cm}}l@{\hspace{0.2cm}}l@{\hspace{0.2cm}}l@{\hspace{0.3cm}}l@{\hspace{0.2cm}}l@{\hspace{0.2cm}}l}
\hline
\hline
 & \multicolumn{3}{l}{ACIS-I} & \multicolumn{3}{l}{ACIS-S} \\
                        & 0 & 1 & 2 & 0 & 1 & 2 \\
\hline
EIS              & 0.43  & 0.36  & 0.15  & 0.86  & 0.13  & 9.4\,10$^{-3}$   \\
members                  & 0.86  & 0.13  & 0.010 & 0.996  & 4.4\,10$^{-3}$   & 9.4\,10$^{-6}$  \\
binaries         & 0.97  & 0.031  & 4.9\,10$^{-4}$  & --  & --  & --  \\
ROSAT            & 0.17  & 0.30  & 0.27  & 0.91  & 0.082  & 3.5\,10$^{-3}$  \\
\hline
\end{tabular}
\end{table}

\section{Analyses} \label{ana}

\subsection{Spectral fits} \label{flux}

To obtain more information about the nature of the X-rays, we have
fitted the spectra of the ten brightest sources with more than 250
counts in the total energy band. We used the CIAO script {\em
psextract} to create spectra from the level-2 events files, and to
construct instrument response files. Data are grouped in bins
containing at least 25 counts. The auxiliary response files are
corrected for the continuous degradation since launch of the ACIS
quantum efficiency at mainly low energies. This is done with the {\em
acisabs}-model by \citet{chargetm} for the time-dependent absorption
due to molecular contaminants on ACIS.  The spectral fitting is done
with {\em Sherpa}, CIAO's package for modeling and fitting data.
Except for CX\,7 on S\,3 (a back-side illuminated CCD with a higher
background countrate than the front-illuminated CCDs), the
contribution from the background was ignored. Energies below 0.35 keV
were excluded from the fits because of the large uncertainty in the
ACIS gain at these energies.

\subsubsection{Late-type stars}
The counterparts of CX\,1, CX\,3, CX\,5, CX\,6, CX\,7, CX\,9 and
CX\,10 are late-type stars.  All are cluster members and binaries,
except CX\,7 (S\,1601) that has not been investigated for binarity and
to which \citet{giraea} assigned a low probability for cluster
membership ($p=48$\,\%, see Table~\ref{optid}), although we note that
\citet{san} found $p=91$\,\% for this star.  We fitted the spectra of
these sources with the {\em xsmekal}-model for emission by an
optically thin gas in collisional equilibrium \citep{meweea}, often
used for the coronal X-ray emission of active late-type stars (see
e.g.\,\citealt{dempea2}). We use the {\em xswabs}-model for
photo-electric absorption to take into account extinction by neutral
hydrogen.  The reddening of M\,67 ($E(B-V)=0.04$) is converted into a
neutral hydrogen column density $N_{\rm H}=2.2~10^{20}$ cm$^{-2}$
following \citet{predschm}; $N_{\rm H}$ is kept fixed in the
fitting. For CX\,7, the possible non-member, this $N_{\rm H}$ is an
upper limit. The total galactic extinction in the direction of M\,67
is equal to the extinction of M\,67; therefore, only when CX\,7 is a
foreground star, it can have an $N_{\rm H}$ that is different
(viz.\,lower) than the one adopted.  For all sources, fits with
one-temperature (1T) solar-abundance models gave results with the
reduced chi-square $\chi^2_{\nu}>2$.  For the brightest sources CX\,1
and CX\,3, we tried to fit two-temperature (2T) models with solar
abundance, and 1T models with variable abundance. While the former
give acceptable fits ($\chi^2_{\nu}<1.4$) with a significant
($>$99\,\%) improvement relative to the 1T solar-abundance models, the
latter do not ($\chi^2_{\nu}>1.8$). For consistency with the brighter
sources, we also fitted the fainter sources with 2T solar-abundance
models. Table~\ref{spec_fits} summarizes the results; for CX\,7 the
fit for $N_{\rm H}=0$ is also included.  The last-but-one column of
the table gives a countrate-to-flux conversion factor $CF$ that
illustrates the intrinsic differences between the spectral models; it
is computed by dividing the unabsorbed flux by the countrate (see
Table~\ref{srclist}). The top frame of Fig.~\ref{spec} shows the best
fit to CX\,1.

We add a few remarks on the spectrum of CX\,1 which is the binary
S\,1063 with an anomalous position in the M\,67 colour-magnitude
diagram, $\sim$1 mag below the giant branch (see \citet{mathea} for a
detailed discussion). Mathieu et al.~note that the colours of
CX\,1/S\,1063 suggest that its reddening is about 5 times higher than
the total reddening in the direction of M\,67, in other words $N_{\rm
H} \approx 1~10^{21}$ cm$^{-2}$. We have looked for indications for
such enhanced, intrinsic reddening in the spectrum of CX\,1 by doing a
fit in which the value of $N_{\rm H}$ is allowed to float. We find
that our data cannot sufficiently constrain $N_{\rm H}$ ($N_{\rm
H}=5^{+8}_{-5}~10^{20}$ cm$^{-2}$; see Table~\ref{spec_fits} for the
complete results) to draw any conclusions. As the countrate of CX\,1
is variable (see Sect.~\ref{var} and Fig.~\ref{lightc}), we also
investigate the possibility of spectral changes by simultaneously
fitting the spectra of the first nine and the last four hours. Fits
for which the temperatures are fixed but for which $N_{\rm H}$ can
vary between the start and end of the observation, give bad results
($\chi^2_{\nu} = 3.2$, 24 degrees of freedom (d.o.f)). Fits for which
$N_{\rm H}$ is fixed to the M\,67 value but for which $kT_2$ can vary,
give better results ($\chi^2_{\nu} = 1.25$, 24 d.o.f.;
$kT_1=0.73_{-0.08}^{+0.09}$ keV, $kT_{\rm 2, start}=2.5_{-0.3}^{+0.5}$
keV, $kT_{\rm 2, end}=4.9_{-0.9}^{+1.3}$ keV).  In the remainder we
use the results of the 2T fit of the single spectrum, with $N_{\rm H}$
fixed to the value of M\,67.

\subsubsection{CX\,2, CX\,4 and CX\,8}
\begin{figure}
\resizebox{\hsize}{!}{\includegraphics{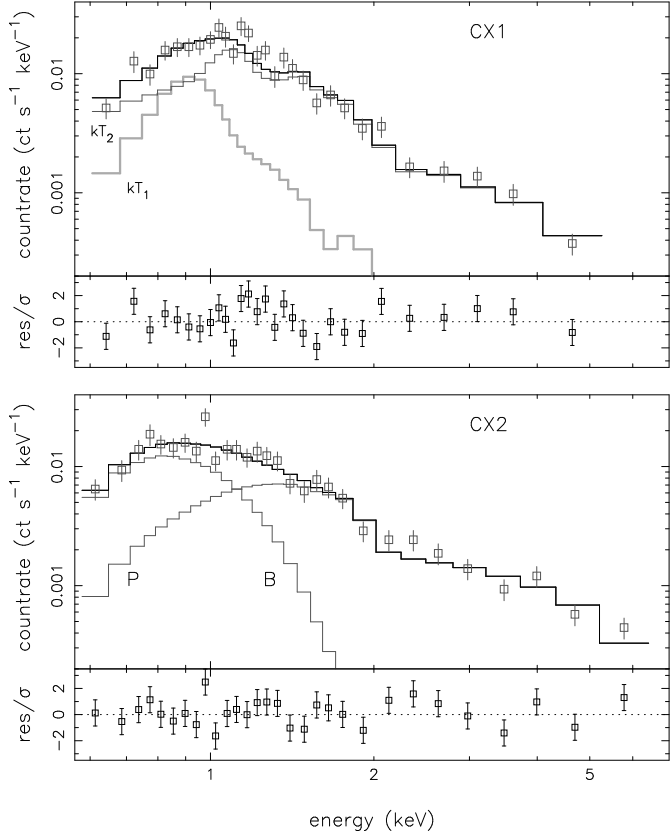}}  
\caption{Spectral fits for the brightest sources. The full lines show
the best fits, squares indicate the data. For CX\,1, the 2T MeKaL
model is decomposed into the cool ($kT_1$; thick line) and hot
($kT_2$) components. For CX\,2, the power law (P) and blackbody
(B) are shown separately.  The lower panels show the residuals of the
fit, normalized to the errors.}
\label{spec}
\end{figure}

For CX\,2, we only know that the optical counterpart is faint and
blue. For CX\,4 and CX\,8 we found faint optical sources near the
X-ray positions in the EIS image. We tried to fit the spectra of
these sources with a power-law, a blackbody, and a 1T-MeKaL model, all
multiplied with a variable $N_{\rm H}$. The power-law and 1T models
gave the best results, but the temperatures in the latter models are
not well constrained and are higher than expected for stellar coronal
sources. The spectra of CX\,2 and CX\,4 show excess soft counts with
respect to the fits, which could be modeled by adding a blackbody to
the power-law and 1T components, or (for CX\,2) by fitting a 2T
model. An F-test indicates that the improvements with respect to the
single-component models are more than 99\,\% significant for CX\,2,
and $\sim$97\,\% significant for CX\,4. The normalizations of the
blackbody-components for CX\,4 are very uncertain. See
Table~\ref{spec_fits} for a summary of the best fits, and the
bottom frame of Fig.~\ref{spec} for the power-law plus blackbody fit
to CX\,2.

\begin{table*}
\caption{Results of spectral fits for the late-type stars (top)
and the remaining sources (bottom). For each source, we give the
best-fitting model (1T and 2T for one- and two-temperature MeKaL
models; P for a power law; B for a blackbody). Next, we give the
temperatures of the MeKaL components $kT_1$ and $kT_2$ (in keV), and
the ratio of their emission measures $EM_1$/$EM_2$ in the case of a 2T
model; the blackbody temperature $kT_{\rm B}$ (in keV) and
normalization $A_{\rm B}=9.9~10^{31} \times (R/d)^2$ with $R$ and $d$
the radius of the emitter and the distance, respectively; the
power-law photon index $\Gamma$; and the column density $N_{\rm H}$
(in cm$^{-2}$). If not mentioned explicitly, $N_{\rm H}$ is kept fixed
at 2.2\,10$^{20}$ cm$^{-2}$. The last three columns give 
the unabsorbed flux $F$ (in erg cm$^{-2}$ s$^{-1}$) and the
countrate-to-flux conversion factor $CF$ (in erg cm$^{-2}$ s$^{-1}$
per ct s$^{-1}$), both in the 0.3--7 keV band, and the reduced
chi-square $\chi^2_{\nu}$/number of degrees of freedom.  Errors are
1-$\sigma$ errors.}

\label{spec_fits}
\begin{tabular}{r@{\hspace{0.13cm}}lllllllllll}

\hline
\hline
CX & mod & $kT_1$ & $kT_2$ & $EM_1$/$EM_2$  & $kT_B$  & $A_{\rm B}$ & $\Gamma$ & $N_{\rm H}$ & $F$ & $CF$ & $\chi^2_{\nu}$/dof \\
   &     &        &        &  & &  &          & {\tiny $\times 10^{21}$} &  {\tiny $\times 10^{-13}$} & {\tiny $\times 10^{-12}$} & \\ 
\hline
 1 & 2T    & 0.79$_{-0.09}^{+0.08}$ & 3.4$_{-0.4}^{+0.4}$ & 0.10 & \multicolumn{1}{l}{--} & \multicolumn{1}{l}{--} & \multicolumn{1}{l}{--} & \multicolumn{1}{l}{--} & 1.59 & 7.7 & 1.32/25  \\
   & 2T    & 0.7$_{-0.1}^{+0.1}$    & 3.2$_{-0.4}^{+0.4}$ & 0.11 & \multicolumn{1}{l}{--} & \multicolumn{1}{l}{--} & \multicolumn{1}{l}{--} & 0.5$_{-0.5}^{+0.8}$    & 1.65 & 8.0 & 1.35/24  \\

 3 & 2T    & 0.86$_{-0.06}^{+0.09}$ & 3.2$_{-0.4}^{+0.5}$ & 0.19 & \multicolumn{1}{l}{--} & \multicolumn{1}{l}{--} & \multicolumn{1}{l}{--} & \multicolumn{1}{l}{--} & 1.06 & 7.4 & 0.74/17  \\ 
 5 & 2T    & 0.86$_{-0.09}^{+0.13}$ & 2.8$_{-0.6}^{+1.2}$ & 0.25 & \multicolumn{1}{l}{--} & \multicolumn{1}{l}{--} & \multicolumn{1}{l}{--} & \multicolumn{1}{l}{--} & 0.75 & 7.4 & 0.48/10   \\
 6 & 2T    & 0.6$_{-0.1}^{+0.1}$    & 1.7$_{-0.3}^{+0.6}$ & 0.52 & \multicolumn{1}{l}{--} &\multicolumn{1}{l}{--} & \multicolumn{1}{l}{--} & \multicolumn{1}{l}{--} & 0.41 & 6.4 & 0.77/5   \\
 7 & 2T    & 0.25$_{-0.05}^{+0.06}$ & 1.09$_{-0.07}^{+0.22}$ & 0.39 & \multicolumn{1}{l}{--} &\multicolumn{1}{l}{--} &  \multicolumn{1}{l}{--} & \multicolumn{1}{l}{--} & 0.32 & 4.7 & 0.71/10  \\
   & 2T    & 0.81$_{-0.1}^{+0.06}$  & 2.3$_{-0.7}^{+1.7}$ & 0.43 & \multicolumn{1}{l}{--} &\multicolumn{1}{l}{--} & \multicolumn{1}{l}{--} & \multicolumn{1}{l}{0} & 0.35 & 5.1 & 0.69/10  \\
 9 & 2T    & 0.8$_{-0.1}^{+0.1}$    & 2.2$_{-0.5}^{+0.8}$  & 0.34 & \multicolumn{1}{l}{--}  &\multicolumn{1}{l}{--} & \multicolumn{1}{l}{--} & \multicolumn{1}{l}{--} & 0.46 & 7.0 & 1.39/4   \\
10 & 2T    & 0.56$_{-0.08}^{+0.05}$ & 2.9$_{-0.8}^{+2.6}$ & 0.53 & \multicolumn{1}{l}{--} & \multicolumn{1}{l}{--} & \multicolumn{1}{l}{--} & \multicolumn{1}{l}{--} & 0.48 & 7.0   & 0.59/4   \\
\hline
 2 & B+P   & \multicolumn{1}{l}{--} & \multicolumn{1}{l}{--} & \multicolumn{1}{l}{--} & 0.12$_{-0.02}^{+0.03}$  & 0.4$_{-0.3}^{2.3}$ & 1.8$_{-0.2}^{+0.2}$ & 3$_{-2}^{+2}$ & 6.70 & 31 & 1.15/25  \\
   & 2T    & 0.53$_{-0.08}^{+0.09}$  & 8$_{-2}^{+3}$  & 0.10 & \multicolumn{1}{l}{--} & \multicolumn{1}{l}{--} & \multicolumn{1}{l}{--}  & $<0.6$ & 1.88 & 8.8 & 1.34/25   \\
   & 1T+B  & 9$_{-3}^{+9}$ & \multicolumn{1}{l}{--}  & \multicolumn{1}{l}{--} & 0.13$_{-0.02}^{+0.03}$ & 0.11$_{-0.09}^{+0.49}$ & \multicolumn{1}{l}{--} & 3$_{-1}^{+2}$  & 4.30 & 20 & 1.20/25   \\
 4 & B+P   & \multicolumn{1}{l}{--} & \multicolumn{1}{l}{--} & \multicolumn{1}{l}{--} & 0.11$_{-0.02}^{+0.02}$ & 3.8$_{-3.5}^{+45.7}$ & 1.7$_{-0.3}^{+0.3}$ & 10$_{-3}^{+3}$ & 33.3 & 230 & 0.97/15   \\
   & 1T+B  & 7$_{-2}^{+23}$ & \multicolumn{1}{l}{--}  &\multicolumn{1}{l}{--} & 0.11$_{-0.02}^{+0.03}$ & 2.4$_{-2.3}^{+20.7}$& \multicolumn{1}{l}{--} & 10$_{-4}^{+3}$ & 25.2 & 174 & 1.04/15   \\
 8 & P     & \multicolumn{1}{l}{--} & \multicolumn{1}{l}{--} & \multicolumn{1}{l}{--} & \multicolumn{1}{l}{--}& \multicolumn{1}{l}{--} & 1.6$_{-0.3}^{+0.3}$ & 4$_{-3}^{+3}$ & 1.12 & 16 & 0.40/7    \\
   & 1T    &  10$_{-5}^{+34}$ & \multicolumn{1}{l}{--}& \multicolumn{1}{l}{--} & \multicolumn{1}{l}{--} & \multicolumn{1}{l}{--} & \multicolumn{1}{l}{--}&  3$_{-2}^{+3}$ & 1.06 & 15 & 0.39/7   \\
\hline
\end{tabular}
\end{table*}

\subsection{Hardness ratios and luminosities} \label{hard}

\begin{figure}
\resizebox{\hsize}{!}{\includegraphics{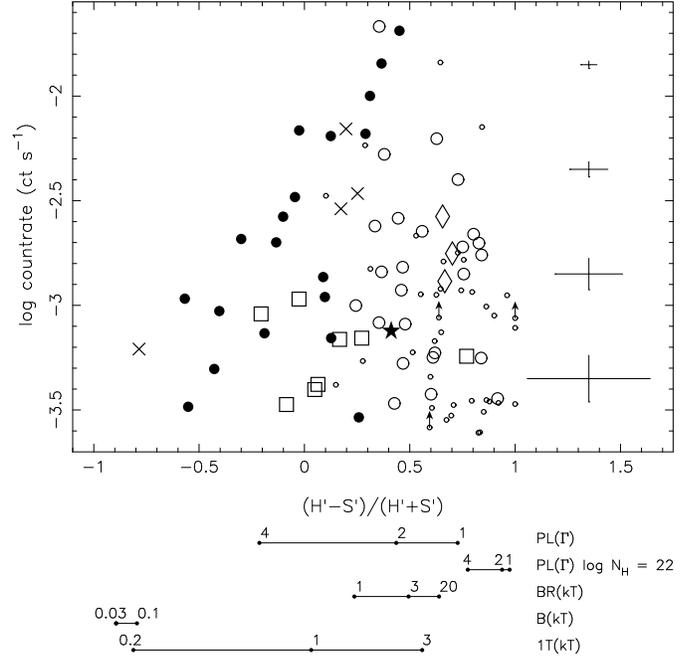}}  
\caption{Hardness ratio $HR$ versus countrate in the total band;
arrows indicated sources only detected in the soft/hard band. Symbols
are as in Fig.~\ref{cmd}; in addition, sources with a faint
counterpart in Table~\ref{fc}, or with a faint counterpart in
Table~\ref{optid-2} without a $B$--$V$, or without an optical
counterpart are indicated with small circles. Typical errors for
different count\-rate levels are plotted on the right. We show the
expected $HR$ for different spectral models assuming $N_{\rm
H}$=$2.2~10^{20}$ cm$^{-2}$: a power law (PL) with $\Gamma$=1,\,2 and
4; thermal bremsstrahlung (BR) with $kT$=1,\,3 and 20 keV; a blackbody
(B) with $kT$=0.03 and 0.1 keV; a 1T MeKaL model (1T) with $kT$=0.2, 1
and 3 keV. For the power law, we show the effect of enhanced reddening
($N_{\rm H}$=$1~10^{22}$ cm$^{-2}$).}
\label{hr}
\end{figure}

For sources without enough counts for a spectral fit, we can obtain
spectral information by comparing the number of counts in different
energy bands. To this end, we compute a hardness ratio
$HR=(H'-S')/(H'+S')$, with $S'$ and $H'$ the counts between 0.5--1 and
1--7 keV, respectively, inside the source regions of the detections in
the total band. Sources with soft spectra have $HR<0$, while for hard
sources $HR>0$. The counts are corrected for vignetting with exposure
maps, and for background. For sources without counts between
0.5--1 keV, $HR \equiv 1$. If $\sigma_{\rm HR} \leq 0.4$, $HR$ is
included in Table~\ref{srclist}. Fig.~\ref{hr} shows $HR$ versus
count\-rate in the total band. Arrows at $HR<1$ indicate sources only
detected in the soft band (0.3--2 keV); as this band partially
includes the energy band used to estimate $H'$, we can still compute
$HR$ for these two sources.  Known (non-)members of M\,67, and stars
near the faint main sequence are mostly found on the ``soft'' side of
the diagram, while the active galaxies, the cataclysmic variable and
sources with faint blue or without counterparts tend to be harder. To
give some idea of what hardness ratios should be expected for
different spectra, we compute $HR$ for: power laws (often used to
describe the emission of ac\-tive galaxies, typically the photon index
$\Gamma=2$); thermal bremsstrahlung and soft blackbodies (for the
spectra of stellar X-ray sources like cataclysmic variables); MeKaL
models (for coronal emission of active stars). We assume $N_{\rm
H}=2.2~10^{20}$ cm$^{-2}$; for the power law, we also show the effect
of enhanced reddening by internal extinction.

For the brightest M\,67 members, unabsorbed X-ray luminosities $L_{\rm
X}$ (0.3--7 keV) are computed with the count\-rate-to-flux conversion
factors $CF$ of Table~\ref{spec_fits}. We find that $CF$ depends on
$HR$: $CF=(4.3(7) \cdot HR + 6(2))~10^{-12}$ erg cm$^{-2}$ s$^{-1}$
per ct s$^{-1}$. This result is based on the spectral fits for the
coronal sources in M\,67 (except CX\,10); as these sources are
relatively hard, extrapolation to softer spectra may give
systematic errors. CX\,10 deviates from this fit, possibly because it
is a blend of two sources (see Sect.~\ref{detect}). For members for
which we did not perform spectral fits with $\sigma_{HR} \leq 0.4$ we
use this relation to compute $L_{\rm X}$; if $\sigma_{HR}>0.4$, we use
the average $HR$ of members to compute a typical $CF$ ($5.6~10^{-12}$
erg cm$^{-2}$ s$^{-1}$ per ct s$^{-1}$). We assume a distance to M\,67
of 820 pc; the results are included in Table~\ref{optid}.

\subsection{X-ray variability} \label{var}

\begin{figure}
\resizebox{\hsize}{!}{\includegraphics[clip]{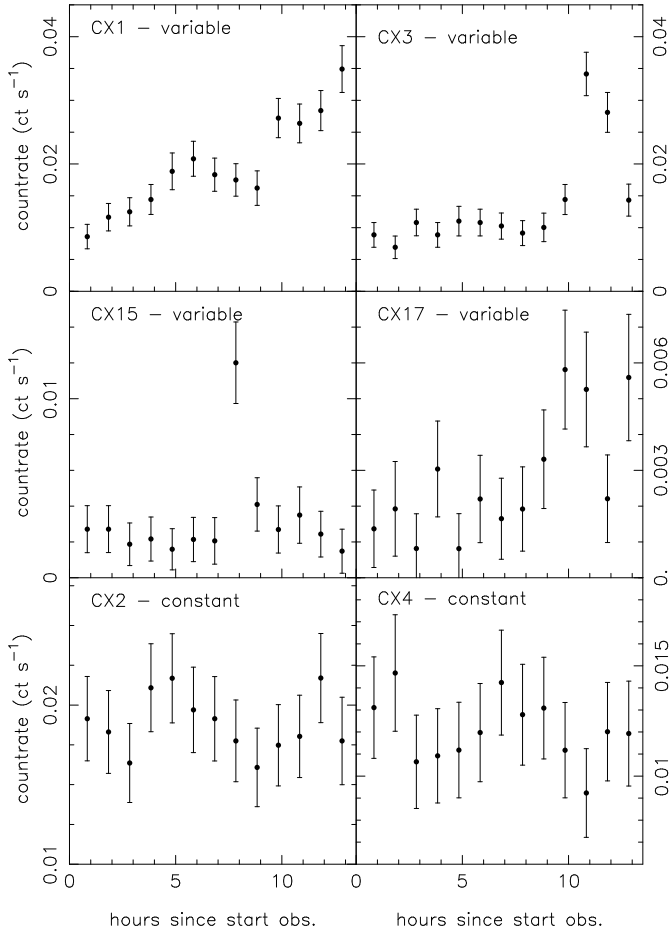}}
\caption{Lightcurves (0.3--7 keV) of the variables and two bright
non-variables. CX\,1 (S\,1063) and CX\,3 (S\,1082) are cluster
members, CX\,15 (S\,972) and CX\,17 (S\,1042) are probable
non-members, and the membership of CX\,2 and CX\,4 is unknown.}
\label{lightc}
\end{figure}

With the CIAO-task {\em lightcurve}, we created light curves in the
soft, hard and total band to look for brightness variations of the
X-ray sources within this 13-hour Chandra observation. Counts were
binned into 1-hour time intervals. The background countrate is
determined for each CCD separately from a region on the chip where no
sources are detected.  The resulting light curves are tested for
variability using a $\chi^2$-test. CX\,1 (S\,1063) and CX\,3 (S\,1082)
are thus found to be variable: the probabilities that their countrates
are constant, are $<$0.4\,\% in all energy bands.  The light curves of
the probable non-members CX\,15 (S\,972) and CX\,17 (S\,1042) have
probabilities $<$4\,\% to be constant, in the total band. Their light
curves are shown in Fig.~\ref{lightc}; for comparison we also show the
lightcurves of the bright non-variables CX\,2 and CX\,4. During the
rise in countrate of CX\,3, there appears to be a small increase in
spectral hardness.  $HR$ averaged over the first nine hours is
$0.23\pm0.06$, while over the last four hours it is $0.49\pm0.06$.
The hardening of CX\,1 is only marginally significant
($HR=0.40\pm0.05$ over the first nine hours while $HR=0.52\pm0.05$ for
the last four hours), while for CX\,15 the hardness ratios at the one
bright point and averaged over the rest of the observation are
compatible ($HR=0.5\pm0.2$ and $0.2\pm0.1$, respectively).  The light
curve of CX\,17 shows no significant spectral changes.

The variations in the light curves of CX\,3 and CX\,15, and the
spectral hardening observed for at least CX\,3, resemble the
behaviour of a flare. In RS\,CVn binaries, flares are frequently
observed. The possible flares in CX\,3 and CX\,15 are poorly
sampled; for CX\,3, the rise time is 1--2 h, while the decay time
could be 2 h or more. These values are not uncommon for RS\,CVns. Also
the peak luminosities, $L_{X} \approx 2$ and $1~10^{31}$ erg s$^{-1}$
(versus ``quiescent'' luminosities of 5~10$^{30}$ and 2~10$^{30}$ erg
s$^{-1}$) of CX\,3 and CX\,15, respectively, are typical, see
e.g.~\citet{franea}.

The spectral fits of Sect.~\ref{flux} suggest that the
variability of CX\,1 is caused by a rise in temperature. This is as
expected for coronal sources for which a flux increase usually
corresponds to an increase in temperature of the coronal plasma.  It
is not likely that a variation in $N_{\rm H}$ causes the variability.
In a point source, less absorption tends to soften the spectrum and,
if anything, the spectrum of CX\,1 at higher countrates is harder.
It could be that the countrate/spectral variability of CX\,1 is
caused by a flare; flare rise times longer than the time span of this
observation have been observed in RS\,CVn systems \citep{ostebrow}.
CX\,1 also shows optical brightness variations that appear to be
quasi-periodic on a period near 23 d \citep[van den Berg et al., in
prep.;][]{vdbergea2002a} that are however not yet understood. The
observed full amplitude $\Delta V \approx 0.15-0.20$ mag, which
corresponds to a flux increase in $V$ of a factor $\approx$1.2. The
variation observed in this Chandra observation is larger, viz.~a
factor $\approx$4 difference between the start and end of the
observation.  Since the optical variability is not periodic, and since
no simultaneous optical and X-ray data are available, it is not
possible to phase the X-ray and optical variations. CX\,1 was also
found to be variable between the two ROSAT observations by a factor
$\approx$1.7 \citep{bellea}.

\section{Results} \label{disc}

We now discuss individual X-ray sources in more detail. In
Sect.~\ref{back}, we estimate the number of background objects among
our sources.  Sect.~\ref{rosat_src} deals with sources already
detected by ROSAT, while new sources are discussed in
Sect.~\ref{new_src}.

\subsection{Background sources} \label{back}

\begin{figure}
\resizebox{\hsize}{!}{\includegraphics{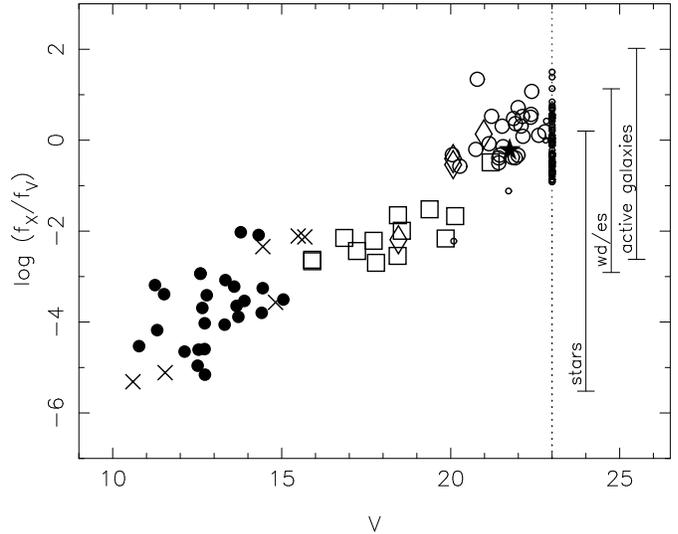}}
\caption{X-ray (0.1--2.4 keV) to optical flux ratio $f_{\rm X}/f_{\rm
V}$ versus $V$ magnitude. Symbols are as in Fig.~\ref{hr}. The ranges
of $f_{\rm X}/f_{\rm V}$ expected for stars (the highest ratios are
for K(e) and M(e) stars in flaring states), white dwarfs (wd) and hot
emission-line stars (es) including cataclysmic variables, and active
galaxies are indicated on the right \protect{\citep{krauea}}. For
sources without optical counterparts we indicate a lower limit to
$f_{\rm X}/f_{\rm V}$ assuming $V=23$ which is the approximate limit
of the EIS image (dotted line). For comparison with the ROSAT sources
from Krautter et al., we have converted our 0.3--7 keV fluxes to the
0.1--2.4 keV band using the fits of Sect.~\ref{flux} for CX\,1--10.
For the other sources, we used a power law with $\Gamma$=2 (for faint
blue sources and galaxies), and an average conversion factor based on
the coronal fits from Sect.~\ref{flux} (for the remaining sources). }
\label{fxfv}
\end{figure}

We use the $\log N - \log S$ distribution of X-ray background sources
as derived from the Chandra Deep Field South \citep{campea,moreea} to
estimate that, between 0.5--2 keV, we expect to detect $\sim$40
extragalactic sources within 8\arcmin~of the ACIS aimpoint
(A. Moretti, priv.~comm.). In the corresponding field, we actually
detect 97 sources between 0.3--2 keV: 28 members and non-members, 9
sources near the lower main sequence, 24 faint blue sources, and 36
sources without counterparts in any of the catalogues searched. We
expect that most background objects are in the last two groups; in
fact, 3 faint blue sources are active galaxies and some faint
counterparts of Table~\ref{fc} look extended. Fig.~\ref{hr} shows that
most sources in these groups have relatively hard spectra as expected
for active galaxies, although we cannot exclude that some are
spectrally hard galactic sources. Similarly, the ratios of their X-ray
and optical fluxes are consistent with those of active galaxies, white
dwarfs or cataclysmic variables, and very active K or M dwarfs (see
Fig.\,\ref{fxfv}).

\subsection{Known X-ray sources} \label{rosat_src}

Here we discuss known X-ray sources for which we can provide new
information. Of the 25 cluster members detected by \citet{bellea},
sixteen were re-detected, and five were outside the ACIS field
(S\,628, S\,364, S\,1112, S\,1113, and the hot white dwarf that was
located between CCDs I\,0 and I\,2). In addition, CX\,21/X\,49,
CX\,20/X\,43, CX\,68/X\,44 and CX\,111/X\,46, that were
identified with the wide binary S\,760, the single members S\,1270 and
S\,775, and the member S\,1027, are now identified with a faint blue
source, with a galaxy, with the non-member S\,2214 and with the member
S\,1024, respectively.

\subsubsection{RS\,CVn binaries}
All active circular binaries with $P_{\rm orb}<10$ d, and the active
eccentric binary S\,1242 \citep{vdbergea}, were re-detected by
Chandra. \citet{bellea} suggested that these sources are similar to
the RS\,CVn binaries, and that their X-rays result from rapid rotation
induced by strong tidal forces. Preliminary orbital periods of $\sim$1
d were recently derived for S\,1019 (CX\,5) and S\,1077 (CX\,10),
which suggests they are also RS\,CVn binaries (D.~Latham,
priv.~comm.). The spectral fits for S\,999, S\,1019 and S\,1077 give
temperatures similar to those typically found for field RS\,CVns using
ROSAT (0.1--2.4 keV) data, viz.~a hot and cool component of
$kT\approx2.2$ and 0.3 keV \citep{dempea2}. The low-energy cutoff of
our data (0.35 keV) may explain the temperature difference of the cool
components. We find a new counterpart for X\,46/CX\,111, viz.~S\,1024
with $P_{\rm orb}$=7.2 d; the former counterpart was S\,1027. S\,1024 is
a circular double-lined spectroscopic binary composed of two nearly
identical stars of $M=1.18$ $M_{\odot}$ and $V=13.4$ that could both
contribute to the X-rays \citep{mathlathea}.

\subsubsection{Peculiar X-ray sources} \label{pec}
\citet{bellea} detected several bright members for which the X-ray
emission is not well understood, viz.~the blue and yellow stragglers
S\,1040 (a giant and a cool white dwarf), S\,1072 and S\,1237 with
orbits that are too wide for strong tidal interaction; the blue
straggler S\,1082; and the eccentric binary S\,1063 with a
sub-subgiant that lies $\sim$1 mag below the giant branch. 

With the results of Sects.~\ref{flux} and \ref{var} we can say
more on the X-rays of three systems. The close binary that was
recently spectroscopically identified in S\,1082
\citep{vdbergea2001ad} could be an RS\,CVn system and thus be the
source of the X-ray emission. The X-ray spectrum, and the variability
in the light curve, are indeed consistent with coronal activity. The
spectra of S\,1040 and S\,1063, for which chromospheric activity
features were found by \citet{vdbergea}, can be fitted by coronal
models as well. However, the nature of the variability of S\,1063
remains unclear.

\subsubsection{Non-members and possible members} \label{nonm}
\begin{figure}
\resizebox{\hsize}{!}{\includegraphics{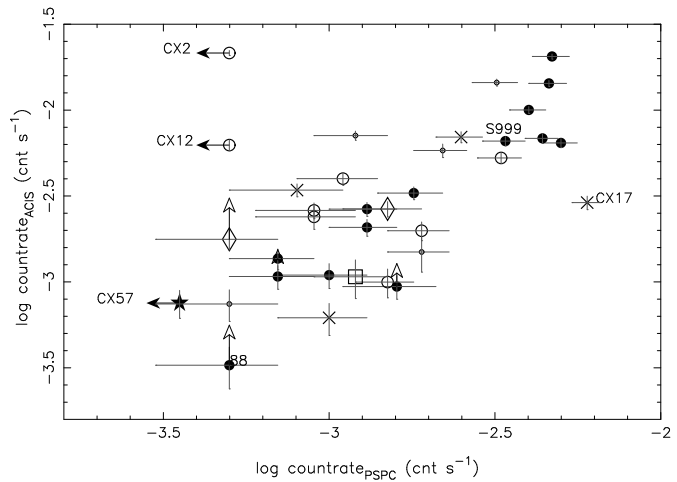}}
\caption{PSPC (0.4-2.4 keV) versus ACIS (0.3-7 keV) count\-rates of
known X-ray sources. Symbols are as in Fig.~\ref{hr}. For ROSAT
sources that are identified with two Chandra sources, we plot the
Chandra countrate of the brightest source, while a vertical arrow
indicates the sum of both countrates. The two new Chandra sources and
possible variables CX\,2 and CX\,12, and the cataclysmic variable
EU\,Cnc (CX\,57) are also indicated.}
\label{count}
\end{figure}

The probable non-members S\,972 (CX\,15), S\,1042 (CX\,17) and S\,1601
(CX\,7) were re-detected. \citet{bellea} considered the brighter of
two possible counterparts for X\,44/CX\,48 (the single member S\,775)
as the most probable counterpart. We now find that the other star,
S\,2214, is the X-ray source. Probably, these late-type stars are
coronal sources: the spectrum of S\,1601 can be well fitted with a
MeKaL model, and the hardness ratios of the other three are consistent
with coronal emission. S\,972 was recently found to have $P_{\rm orb}
= 1.2$ d (D.~Latham priv.\,comm.), and is thus an RS\,CVn system. The
light curves of S\,972 and S\,1042 show indications of variability
(see Sect.~\ref{var}), possibly due to flares. S\,1042 was also found
to be variable by \citet{bellea}. Fig.~\ref{count}, that compares PSPC
(0.4--2.4 keV) and ACIS (0.3--7 keV) countrates of sources detected by
both instruments, suggests that it also changed brightness between the
present and the ROSAT observations.

\subsubsection{The cataclysmic variable EU\,Cnc}
The cataclysmic variable EU\,Cnc (CX\,57) was detected by ROSAT only
below 0.4 keV. From Fig.~\ref{count} (in which CX\,57 is plotted
below the PSPC detection limit at $\log$ (countrate) = $-3.3$ for
clarity), we indeed estimate that its Chandra countrate corresponds
to an expected PSPC countrate between 0.4--2.4 keV near the detection
limit. We find a relatively high hardness ratio. The combination of a
hard and soft spectral component is often seen in cataclysmic
variables where the white dwarf has a high magnetic field
(i.e.\,polars), as is probably the case in EU\,Cnc; the hard X-rays
are believed to come from shocks in the accretion stream while the
soft X-rays come from the heated surface of the white dwarf \citep[for
referen\-ces, see][]{kuulea}. We refer to \citet{bellea} for more
details on EU\,Cnc. Assuming a thermal-bremstrahlung model, we
estimate that $L_{\rm X} \approx 5~10^{29}$ erg s$^{-1}$ (0.3--7 keV).

\subsection{New X-ray sources} \label{new_src}

Of the 25 proper-motion cluster members detected by Chandra, ten are
new sources that are discussed in Sects.~\ref{newwuma} to
\ref{newnobin}. We discuss sources that we suspect to be close active
binaries in M\,67 in Sect.~\ref{newms}, a new bright and variable
source in Sect.~\ref{newcx2}, and probable non-members in
Sect.~\ref{newnonm}.

\subsubsection{The contact binaries S\,757 and ET\,Cnc} \label{newwuma}
\begin{figure}
\resizebox{\hsize}{!}{\includegraphics{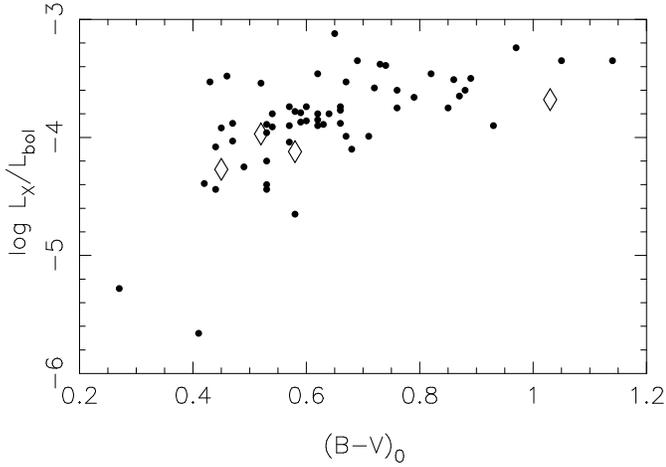}}  
\caption{X-ray luminosity (0.1--2.4 keV) normalized to the bolometric
luminosity versus $(B-V)_0$ for (mainly) field contact binaries
(filled circles) and for those in M\,67 (diamonds). Figure adapted
from \citet{stepea}.}
\label{wuma}
\end{figure}

With the detection of S\,757 (CX\,23) and ET\,Cnc (CX\,61) all four
known contact binaries in M\,67 are now detected in X-rays. While
S\,757 is a proper-motion cluster member, no proper motion has been
measured for the faint system ET\,Cnc. If it is located at the
distance of M\,67, it lies on the $M_V$-colour-period relation for
contact binaries and therefore it probably belongs to the cluster
\citep{ruci}. The X-rays of contact binaries are believed to result
from coronal activity around the rapidly rotating stars in the
binary. Fig.~\ref{wuma} shows that $L_{\rm X}/L_{\rm bol}$ of the
systems in M\,67 is comparable to that of other contact binaries of
similar colours, i.e.~several times lower than the maximum X-ray
emission ($\log L_{\rm X}/L_{\rm bol} \approx -3$) of active single or
detached binary stars (see St{\c e}pie\'n et al.~2001).

\subsubsection{RS\,CVn binaries}
Five new sources are identified with binaries in M\,67 with
$P_{\rm orb} < 12$ d.  The circular orbits of S\,1009 (CX\,78) and
S\,986 (CX\,157) indicate that tidal interaction is taking place; the
X-rays are then expected to be due to rapid synchronous
rotation. S\,1009 ($P_{\rm orb}=5.95$ d) lies $\sim$1 mag below the
main-sequence turnoff \citep{lathmathea}. S\,986 ($P_{\rm orb}=10.3$
d) is also a spectroscopic binary near the turnoff \citep{mathlathea},
that shows eclipses \citep{sandshet986}. The secondary hardly
contributes to the optical flux, but \citet{sandshet986} recently
discovered a third component in the spectrum; it is not clear whether
this star is physically bound to the binary. Their decomposition puts
the primary to the blue of the turnoff, which makes it a possible blue
straggler.  S\,1272 (CX\,155) is an eccentric binary ($e=0.26$) with
a primary that just left the main sequence
\citep{mathlathea}. Possibly, tidal interaction started when the
primary began to expand, and since synchronization is achieved before
circularization, the primary could already be spun up (the
pseudo-synchronous rotation period is 7.9 d \citep{hut}).  S\,996
(CX\,91) and S\,773 (CX\,72) have preliminary orbital periods of
6.7 and 5.7 d, respectively, (D.~Latham, priv.~comm.)  which suggest
they are RS\,CVn binaries, too.  As S\,996 was considered as a
possible counterpart of X\,16 (CX\,57/EU\,Cnc), \citet{pasqbell} took
low-resolution optical spectra of this star, but no activity features
were found.

While our active binaries show a spectral hardening with increasing
brightness (Fig.~\ref{hr}), X-ray sources in the Pleiades (a $\sim$100
Myr open cluster) show an inverse trend. In the Pleiades, this effect
is mainly due to the early spectral types of the X-ray brightest
members, whose soft X-rays may be different in nature
\citep{daniea}. If we disregard these early-type sources, no strong
dependence of countrate and hardness ratio is seen in the Pleiades.

\begin{figure}
\resizebox{\hsize}{!}{\includegraphics{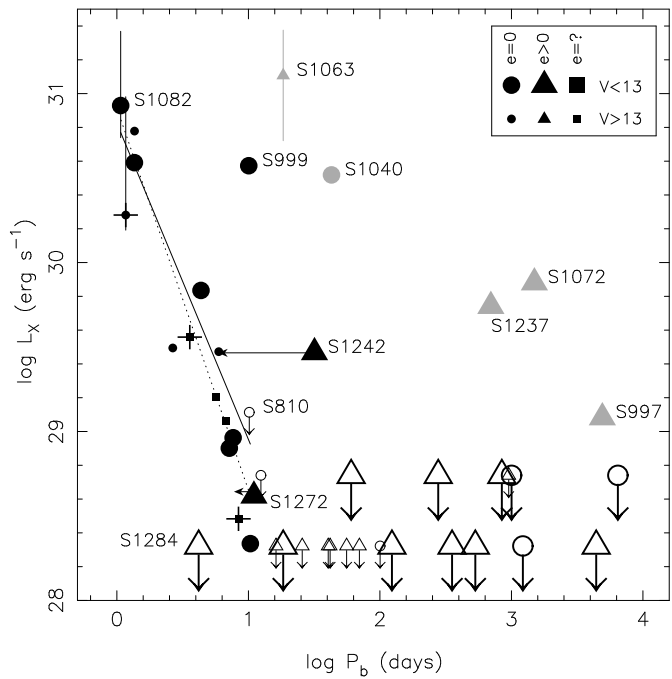}}
\caption{Orbital period versus X-ray luminosity (0.3--7 keV) of M\,67
binaries in the ACIS field. Filled (open) symbols are detections
(upper limits), plusses are detected suspected binaries in M\,67
(CX\,15, CX\,58, CX\,141). The shape of the symbols indicate the
magnitude and eccentricity.  The peculiar binaries of
Sects.~\ref{pec} and \ref{bs} are plotted in gray. Individual
sources discussed in the text are labelled. Upper limits are
estimated by looking for the faintest source as a function of distance
from the aimpoint. We estimate that at 0\arcmin--6\arcmin,
6\arcmin--9\arcmin\, and 9\arcmin--11\arcmin\,from the aimpoint, the
limiting $L_{\rm X}$ for a coronal source in M\,67 is
$\sim$2.1~10$^{28}$, 5.5~10$^{28}$ and 1.3~10$^{29}$ erg
s$^{-1}$. The average $L_{\rm X}$ is plotted for the variables,
but luminosity ranges are indicated. The fits of $\log L_{X}$ to $\log
P_{\rm orb}$ including S\,999 (full line), and excluding S\,999 and
using the ``quiescent'' $L_{\rm X}$ of S\,1082 (dotted line) are
shown.}
\label{binaries}
\end{figure}

Fig.~\ref{binaries} shows the X-ray luminosities of all binaries in
M\,67 versus orbital period; arrows indicate the pseudo-synchronous
rotation periods of S\,1242 and S\,1272 (note that the object
near the arrowhead of S\,1242 is S\,1009).  When studying
activity-rotation relations, $L_{\rm X}$ is often normalized to the
bolometric luminosity.  Since in some of our binaries and triples
more than one star contribute to the X-ray and/or optical flux, we
make no attempt to do this to avoid introducing errors.  However, in
Fig.~\ref{binaries} we do indicate whether a system is brighter or
fainter than the main-sequence turnoff around $V=13$.

Unlike \citet{dempea,dempea3} and \citet{pallea} who found no
correlation between $L_{\rm X}$ and $P_{\rm orb}$ of active binaries
using Einstein and ROSAT All Sky Survey data, respectively, our
data suggest a decrease of $L_{\rm X}$ with increasing $P_{\rm orb}$
for periods $\lesssim$ 12 d for which we expect that the rotation
period equals the orbital period\footnote{A linear fit including
the confirmed M\,67 members but excluding the peculiar systems of
Sects.~\ref{pec} and ~\ref{bs}, and assuming $\sigma_{L_{\rm
X}}$=50\,\%, gives $\log L_{X} \propto -1.9(2) \times \log
P_{\rm orb}$ and a correlation coefficient of --0.75; for S\,1242 and
S\,1272, we assumed pseudo-synchronous rotation. If we exclude
S\,999 and use the quiescent $L_{\rm X}$ of S\,1082, we find $\log
L_{X} \propto -2.3(2) \times \log P_{\rm orb}$ and a correlation
coefficient of --0.95. We note that this relation is similar to the
$\log L_{X} \propto 2 \log (v\sin i)$ relation found for single
late-type stars by \citet{pallea}.}. The difference may be due to the
fact that our sample is homogeneous in age: old enough that all
binaries out to $\sim$12 d are synchronous, and that no stars are
rotating rapidly without tidal interaction. 
\citet{dempea,dempea3} emphasize that their sample of almost 200
active binaries, taken from the Catalogue of Active Binaries
\citep{straea}, is not necessarily homogeneous but instead also
contains systems that may not be true RS\,CVns.

We note that S\,999 is considerably above the observed general trend,
being $\sim$100 times brighter than S\,986, which has similar
photometric and orbital properties. Since Fig.~\ref{count} suggests no
large variation in the countrate of S\,999, a flare may not explain
the relatively high luminosity.

\subsubsection{The eccentric long-period blue straggler S\,997} \label{bs}
S\,997 (CX\,95) is a blue straggler in an eccentric orbit of 4913
d. The primary has $T_{\rm eff} \approx 6600$ K and $v \sin i \approx
20$ km s$^{-1}$ \citep{lathmilo,sandshet}. In normal late-type stars
the combination of X-ray activity and such a high $v \sin i$ is not
unusual, but other M\,67 blue stragglers with similar $T_{\rm eff}$
and $v \sin i \gtrsim 20$ km s$^{-1}$ (S\,975, S\,1195) are not
detected in X-rays. The eccentric orbit suggests that no strong tidal
interaction is taking place; moreover, the orbit is too wide for rapid
synchronous rotation of a possible late-type secondary. Like for the
other detected blue and yellow stragglers with wide eccentric orbits
(S\,1072, S\,1237), we do not understand the X-rays of
S\,997. Possibly, like S\,1082, it contains a close binary that so far
went undetected because it is faint or has spectral lines that are
broadened by rotation.

\subsubsection{Other members} \label{newnobin}
S\,1281 (CX\,94) shows no radial-velocity variations in six
observations spanning 3923 days (D.\,Latham, priv.\,comm.). This star
is marked as a possible blue straggler by \citet{sand}. S\,1050
(CX\,104) shows radial-velocity variations that suggest an orbital
period longer than 12 d (D.\,Latham, priv.\,comm.). If both stars are
indeed single or long-period binaries, their X-rays are still
unexplained.

S\,1203 (CX\,67) has never been investigated for binarity. It lies
somewhat above the main sequence, and could be a photometric binary
\citep{montea}.

\subsubsection{Sources near the main sequence} \label{newms}
\begin{figure}
\resizebox{\hsize}{!}{\includegraphics{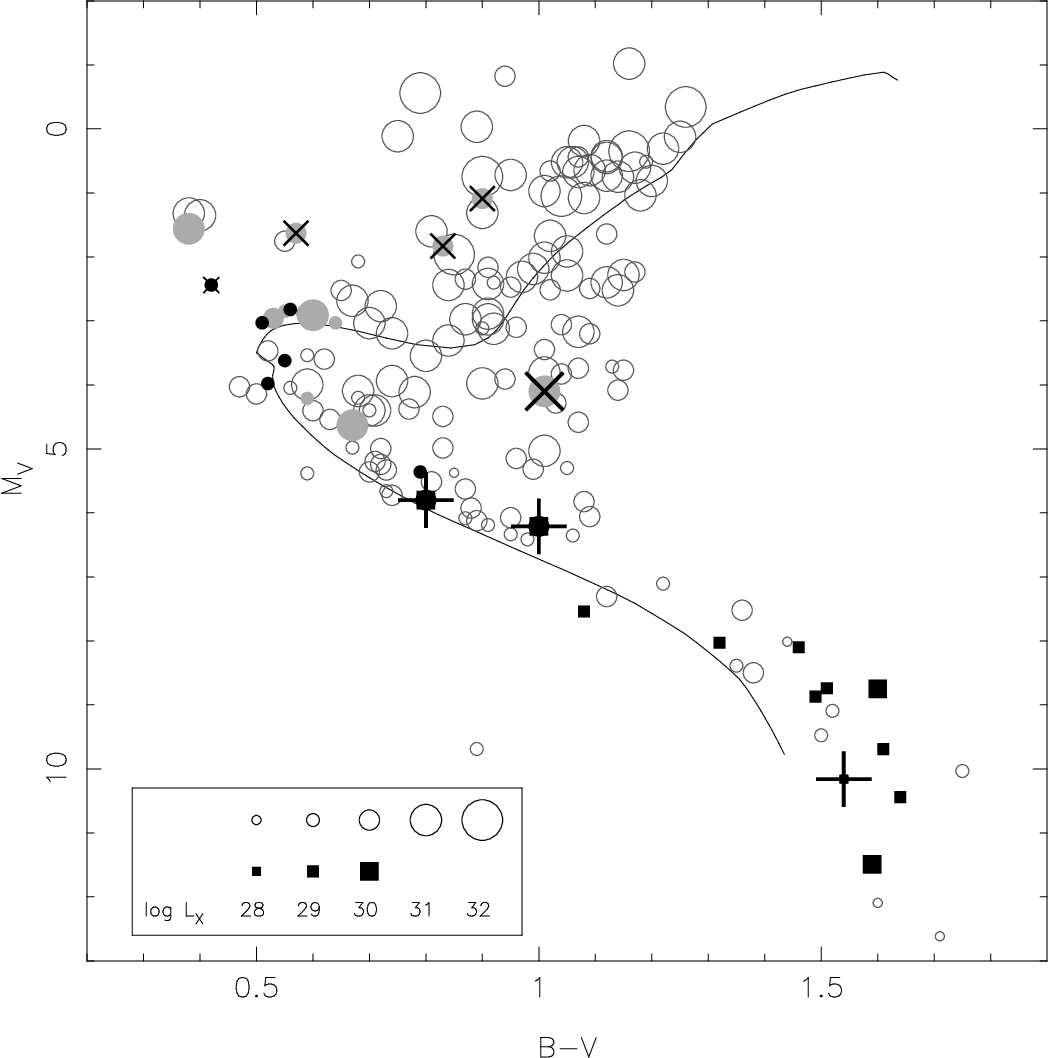}}  
\caption{Colour-magnitude diagram with field RS\,CVns from
\protect{\citet[][open circles]{dempea,dempea3}}. Active binaries in
M\,67 are filled circles; filled squares are the candidate members of
Sect.~\ref{newms}. The peculiar binaries of Sects.~\ref{pec} and
\ref{bs} are shown as large $\times$ symbols, the suspected binaries
in M\,67 (CX\,15, CX\,58, CX\,141) as large $+$ symbols.  Black (gray)
symbols are new (known) X-ray sources. The size of the symbol
indicates the X-ray luminosity. Photometry of M\,67 stars is corrected
for reddening and distance modulus. For reference, we also show an
isochrone with the age of M\,67 \protect{\citep[4 Gyr,][]{polsea}}.}
\label{xcmd}
\end{figure}

The optical counterparts of twelve new sources lie near the lower end
of the M\,67 main sequence. These sources likely belong to M\,67;
large contamination from foreground or background stars is not
expected since M\,67 has a high galactic latitude
($b=32^{\circ}$). The hardness ratios of many are compatible with
coronal emission (see Fig.~\ref{hr}). Fig.~\ref{xcmd}, which is an
extended version of Fig.~4 in \citet{bellea}, shows that if these
sources are cluster members, their X-ray luminosities are comparable
to those of field active binaries with late-type main-sequence
primaries.  The counterparts are too faint to be included in
proper-motion or radial-velocity studies, but for some we have
additional photometric information.  CX\,58 is a star on the binary
sequence, for which \citet{stasea2002} derive a photometric period of
\Pp$=3.58$ d (their star 2703). CX\,141 is EY\,Cnc with \Pp$=8.4$ d
\citep{gillea}. We suggest that these stars are active binaries in
which spots on the surface of the active star cause the brightness
variations. In that picture, the photometric period is the rotation
period, which for synchronous rotation equals the orbital
period. Other stars that \citet{montea} note to be on the binary
sequence are CX\,76, CX\,77 and CX\,153. CX\,76 shows brightness
variations, but no period was found \citep[][ their star
3348]{stasea2002}.

\subsubsection{The bright new source CX\,2/E\,510} \label{newcx2}

CX\,2 is a new bright source with a faint ($V\approx21$) blue optical
counterpart, source 510 in the EIS list. It is not clear whether
E\,510 belongs to M\,67, or is instead a background star or
galaxy. From Fig.~\ref{count} we estimate that Chandra sources with a
countrate comparable to that of CX\,2 would have been easily detected
by ROSAT. We conclude that CX\,2 must have brightened at least an
order of magnitude since the time when the ROSAT observations were
taken (1991, 1993). The spectrum of CX\,2 has a hard and soft
component; the $N_{\rm H}$ is higher than the value for M\,67 and is
therefore likely to be intrinsic. Like EU\,Cnc (CX\,57) it could be a
magnetic cataclysmic variable. If CX\,2 belongs to M\,67, we find $L_X
\approx 5~10^{31}$ erg s$^{-1}$ (0.3--7 keV) which is inside the
observed range for cataclysmic variables. However, we find a blackbody
temperature ($\sim$120 eV) that is on the high side for polars, that
have typical soft components of 30--100 eV. The ratio of the typical
size of the X-ray emission area $R$ and the distance, as derived from
the normalization of the blackbody, is 6.0~10$^{-17}$, which for the
distance of M\,67 is quite small ($R=1.5$ km) for a white dwarf. To
check if CX\,2 could be a binary in M\,67 with a neutron star or a
black hole, we compare its X-ray luminosity with those of low-mass
X-ray binaries in quiescence. These sources are believed to emit
X-rays due to thermal emission from the cooling neutron star surface
and/or due to low-level accretion from a low-mass
secondary. \citet{garcea} give $L_X$ normalized to the Eddington
luminosity $L_{\rm Edd}$, for both neutron-star and black-hole
systems. For CX\,2, we find $\log (L_X/L_{\rm Edd}) \approx -6.6$
assuming CX\,2 contains a 1.4 $M_{\odot}$ neutron star, and $\log
(L_X/L_{\rm Edd}) \approx -7.3$ assuming a 7 $M_{\odot}$ black hole;
the upper limits from the ROSAT detection are ten times
lower. Comparison with Fig.~1 from \citet{garcea} shows that these
luminosity ranges are in agreement with systems containing a black
hole, while all neutron-star systems but one (the milli-second pulsar
SAX\,J1808-365) are brighter. The X-ray spectra of quiescent
black-hole low-mass X-ray binaries can be well fitted by power laws
with $\Gamma \approx 2$ \citep{kongea}, but excess soft emission as we
find for CX\,2 is not reported in the (mostly faint with $<150$
counts) Chandra spectra fitted by \citet{kongea}. If CX\,2 lies beyond
M\,67, $L_X$ increases accordingly allowing the possibility of a
neutron-star system. Note that, in either case, the luminosity
variation between the ROSAT and Chandra observations also needs to be
explained. Clearly, more information is needed to establish the nature
of this system.

We note that we detected CX\,2 in archival XMM-Newton observations
taken on 2001 October 21, almost five months after our Chandra
observation. Preliminary analysis of the EPIC-pn spectrum ($\sim$200
source counts) shows that its flux has decreased to about 75\%\ of the
flux level during the Chandra observations.  Here we assumed the
parameters of the best-fit spectrum in Table~\ref{spec_fits}.

Another possible variable source with a faint counterpart is CX\,12,
that has a hard spectrum (see Fig.~\ref{count}).

\subsubsection{Probable non-members} \label{newnonm}
CX\,109 and CX\,118 are the probable non-members S\,1466 and S\,1013,
respectively, whose photometry places them among the blue stragglers
of M\,67. A deviant proper motion could be due to a close encounter
that created a blue straggler, as suggested for S\,1466 by Landsman et
al.~1998. If they are foreground dwarfs, $L_{\rm X} \approx$
1--2~10$^{28}$ erg s$^{-1}$; if they are background giants, $L_{\rm X}
\approx$ 2--4~$10^{29}$ erg s$^{-1}$. This is inside the observed
range for early-F dwarfs/giants \citep{schm,maggea}. Neither star has
been investigated for binarity. S\,1466 was also detected in
ultraviolet by \citet{landea98}; its 1520\,\AA -- $V$ colour suggests
it is an A\,8 star (W.\,Landsman, priv. comm.), only slightly hotter
than the F\,0-type derived from its $B-V$. \nocite{landea98}

\section{Summary and conclusions} \label{conc}

We detected 25 cluster members, including ten new sources. Most of
these members are binaries, which is not unexpected as old single
stars are faint in X-rays. Almost all binaries with $P_{\rm orb}$$<$
12 d in the ACIS field of view are detected (Fig.~\ref{binaries});
their X-ray luminosities decrease with increasing orbital period. For
detached binaries with main-sequence or subgiant primaries, this
period corresponds to the period where the transition from
circularized to eccentric orbits occurs in M\,67 \citep[$P_{\rm
cut}=12$ d, ][]{lathmathea}.  The detection of most binaries with
$P_{\rm orb} < P_{\rm cut}$ is therefore consistent with the
explanation of their X-rays: the activity is enhanced by rapid
synchronous rotation. Two binaries with $P_{\rm orb}<P_{\rm cut}$ are
not detected. For S\,810 ($P_{\rm orb}$=10.2 d) this could be due to
the high upper limit for $L_{\rm X}$. S\,1284 ($P_{\rm orb}$=4.2 d) is
a binary of a late-A blue straggler while the lower limit on the
secondary mass is compatible with a white dwarf and a late-type dwarf
\citep{milolath92}. The eccentric orbit ($e=0.21$) indicates that
strong tidal interaction is not taking place.

Surprisingly, we detected one cluster member, S\,1281, that shows no
radial-velocity variations. Also the detection of the systems with
$P_{\rm orb} \gg 12$ d, in which strong tidal interaction is not
expected, remains a mystery. Our spectral fits indicate that the
X-rays of the sub-subgiant S\,1063 and the giant plus white dwarf
S\,1040, are coronal.

The twelve X-ray sources near the main sequence are candidates for
being close binaries, but their membership and binary status should be
established by spectroscopy. We also detected a group of sources with
optically faint blue counterparts, including a new bright source that
must have brightened at least ten times since the time of the ROSAT
observations. While we expect that most of them are background
sources, some could be interacting binaries in M\,67.

\begin{acknowledgements}
We thank D.~Latham for sharing unpublished results of the M\,67
radial-velocity study carried out by D.~Latham, A.~Milone, and
R.D.~Mathieu. We also wish to thank A.~Moretti for assistance with the
data reduction and the discussion of Sect.~\ref{back}; and J.~Homan
for discussions on Sect.~\ref{newcx2} and for the analysis of
XMM-Newton data. MvdB acknowledges financial support from the Italian
Space Agency and MIUR.
\end{acknowledgements}

\bibliographystyle{aa}
\bibliography{m67chandra_astroph.bbl}

\clearpage
\newpage

\setcounter{table}{0}
\begin{table*}
\caption{X-ray sources in the field of M\,67 detected by Chandra.  For
each source we list the Chandra number, the distance from the
aimpoint, the celestial position (J\,2000) derived from the image in
the total energy band, the number of background-corrected counts in
the total energy band derived by {\em wavdetect}, the net countrate
(i.e.\,corrected for background counts and for vignetting) in the
total band, the energy band $B$ in which the detection has the highest
significance (S=0.3$-$2 keV, H=2$-$7 keV, T=0.3$-$7 keV), and the
hardness ratio $HR$ (if $\sigma_{HR} < 0.4$). If appropriate, we also
give the ROSAT counterpart 'X' \protect{\citep{bellea}}, the optical
counterpart 'opt', and the distance $d_{\rm XO}$ between the Chandra
and the optical source. The name of the optical counterpart is
selected from (in order of preference): \protect{\citet{san}} (S), the
EIS list of \protect{\citet{momaea}} (E), and \protect{\citet{fanea}}
(F). Sources that are identified by eye with objects listed in neither
catalogue are labelled with 'faint' (see Table~\ref{fc}).}

\begin{tabular}{rrllrcrrlll}
\hline \hline
CX  & $d_{\rm aim}$ & \multicolumn{1}{c}{$^{**}\alpha_{\rm J2000}$}  & \multicolumn{1}{c}{$^{**}\delta_{\rm J2000}$} & counts & ctr & $B$ & \multicolumn{1}{c}{$HR$} & X   & opt  & $d_{\rm XO}$  \\
    & \multicolumn{1}{c}{(\arcmin)}  &  \multicolumn{1}{c}{($^h$,$^m$,$^s$)}  &  \multicolumn{1}{c}{($^{\circ}$,\arcmin,\arcsec)}   &    & (c\,ks$^{-1}$) &  &  &      &      & (\arcsec) \\ \hline
  1 &  3.19   &   8 51 13.373(2)      & 11 51 40.23(2)  & 892(30)  & 20.5 &  T &  0.45$\pm$0.03   &  8   & S\,1063    & 0.27 \\
  2 &  3.78   &   8 51 38.456(1)    &   11 49 06.28(3)  & 872(30)  & 21.5 &  T &  0.36$\pm$0.03   &      & E\,510     & 0.13 \\
  3 &  3.93   &   8 51 20.798(2)    &   11 53 26.40(3)  & 619(25)  & 14.3 &  T &  0.37$\pm$0.04   &  4   & S\,1082    & 0.23 \\
  4 &  7.23   &   8 50 57.104(8)    &   11 46 07.57(8)  & 513(24)  & 14.5 &  T &  0.65$\pm$0.04   & 15$^{c}$ & faint  & 0.1 \\
  5 &  1.10   &   8 51 23.303(1)    &   11 48 26.92(2)  & 446(21)  & 10.0 &  T &  0.31$\pm$0.05    & 11  & S\,1019    & 0.22 \\
  6 &  0.32   &   8 51 23.794(2)    &   11 49 49.56(3)  & 286(17)  & 6.4 &  T &  0.13$\pm$0.07   & 10   & S\,1040$^d$    & 0.07 \\
  7 & 13.20   &   8 52 16.88(3)     &   11 48 32.1(5)   & 282(21)  & 7.0 &  S &  0.20$\pm$0.09    & 47  & S\,1601    & 0.61 \\
  8 &  5.95   &   8 51 27.129(7)    &   11 55 24.97(9)  & 279(17)  & 7.1 &  T &  0.84$\pm$0.04  & 32    & faint      & 0.5 \\
  9 &  2.72   &   8 51 18.715(3)    &   11 47 02.96(3)  & 272(17)  & 6.6 &  T &  0.29$\pm$0.06  & 13    & S\,999     & 0.10 \\
 10 &  5.22   &   8 51 07.217(5)    &   11 53 02.22(6)  & 250(16)  & 6.9 &  S & $-$0.03$\pm$0.07  &  7  & S\,1077    & 0.59 \\
 11 & 12.78   &   8 52 13.78(4)     &   11 46 26.0(6)   & 231(21)  & 5.8 &  S &  0.29$\pm$0.10  & 51    &            &  \\
 12 &  8.18   &   8 51 28.40(1)     &   11 41 28.2(2)   & 225(16)  & 6.3 &  T &  0.63$\pm$0.06  &       & E\,2024    & 0.51 \\
 13 &  5.62   &   8 51 03.818(8)    &   11 46 30.15(9)  & 218(15)  & 5.3 &  T &  0.38$\pm$0.07  & 14$^{c}$ & E\,2153 & 0.35 \\
 14 &  7.46   &   8 51 04.69(1)     &   11 55 29.7(1)   & 148(13)  & 4.0 &  T &  0.73$\pm$0.07  & 34  & E\,2823      & 0.16 \\
 15 &  5.22   &   8 51 17.758(9)    &   11 44 29.5(1)   & 141(12)  & 3.4 &  T &  0.25$\pm$0.09  & 17  & S\,972       & 0.27 \\
 16 &  3.88   &   8 51 37.888(4)    &   11 50 57.31(6)  & 140(12)  & 3.3 &  S & $-$0.05$\pm$0.10  & 40  & S\,1282    & 0.21 \\
 17 &  3.72   &   8 51 08.006(4)    &   11 49 56.29(5)  & 125(11)  & 2.9 &  S &  0.17$\pm$0.10  & 42  & S\,1042      & 0.53 \\
 18 & 15.74   &   8 52 25.11(5)     &   11 45 21(1)     & 122(18)  & 3.3 &  T &  0.10$\pm$0.24  &     & 	      &  \\
 19 &  1.24   &   8 51 28.170(3)    &   11 49 27.76(4)  & 116(11)  & 2.7 &  T & $-$0.10$\pm$0.11  & 45  & S\,1036    & 0.20 \\
 20 &  6.59   &   8 51 49.99(1)     &   11 49 52.9(2)   & 106(11)  & 2.7 &  T &  0.66$\pm$0.09  & 43  & E\,2757  & 0.40 \\
 21 &  7.66   &   8 50 52.72(2)     &   11 47 43.7(2)   & 103(11)  & 2.6 &  T &  0.45$\pm$0.10  & 49  & E\,2183  & 0.66 \\
 22 &  5.65   &   8 51 46.16(1)     &   11 49 50.3(3)   & 92(10)   & 2.3 &  T &  0.56$\pm$0.10  &     & E\,2756  & 0.64 \\
 23 &  5.74   &   8 51 04.86(2)     &   11 45 56.9(2)   & 82(9)    & 2.0 &  S & $-$0.13$\pm$0.13  &     & S\,757 & 0.23 \\
 24 &  3.11   &   8 51 21.78(1)     &   11 52 38.2(2)   & 81(9)    & 2.1 &  S & $-$0.30$\pm$0.13  & 37  & S\,1072$^d$  & 0.11 \\
 25 & 15.08   &   8 52 22.59(5)     &   11 45 37.3(9)   & 80(14)   &  2.2 &  T &  0.53$\pm$0.13  &       &      &      \\
 26 & 13.62   &   8 52 07.37(6)     &   11 41 18(1)     & 79(12)   &  2.4 &  S &  0.34$\pm$0.17  & 55  & E\,2018$^{g}$     & 3.27 \\
 27 &  1.31   &   8 51 22.82(1)     &   11 48 14.2(2)   & 79(9)    &  1.8 &  T &  0.73$\pm$0.10  &     &                 &      \\
 28 &  3.99   &   8 51 24.81(1)     &   11 53 31.3(2)   & 75(9)    &  1.7 &  T &  0.84$\pm$0.09  &     & E\,2297    & 0.23 \\
 29 &  2.19   &   8 51 30.58(1)     &   11 50 44.8(1)   & 71(8)    &  1.6 &  T &  0.66$\pm$0.12  &     &                              &      \\
 30 & 17.87   &   8 52 27.72(7)     &   11 41 14(1)     & 66(14)   &  2.2 &  S &  0.80$\pm$0.15  &     & E\,210$^{f}$	             & 2.76 \\
 31 &  3.44   &   8 51 35.70(1)     &   11 48 00.9(2)   & 66(8)    &  2.0 &  T &  0.83$\pm$0.10  & 48  & E\,2190      & 0.23 \\
 32 &  6.32   &   8 50 58.74(2)     &   11 51 38.8(3)   & 62(8)    &  1.8 &  T &  0.70$\pm$0.10  & 39  & E\,1268$^{h}$ 	      & 1.97 \\
 33 &  6.27   &   8 51 46.61(2)     &   11 52 02.9(3)   & 62(8)    &  1.5 &  T &  0.47$\pm$0.14  &     & E\,600             & 0.21 \\
 34 & 11.78   &   8 52 10.41(5)     &   11 47 21.0(6)   & 58(10)   &  1.4 &  S &  0.37$\pm$0.20  &     & E\,2722$^{f}$	     & 1.56 \\
 35 & 14.59   &   8 51 53.18(6)     &   11 36 56.8(9)   & 56(10)   &  1.9 &  T &  0.75$\pm$0.12  &     & E\,1929$^{f}$      & 2.22 \\
 36 &  4.21   &   8 51 31.29(1)     &   11 45 51.0(2)   & 53(7)    &  1.4 &  S &  0.09$\pm$0.17  & 53  & S\,1234   & 0.37 \\
 37 &  4.60   &   8 51 27.38(2)     &   11 45 04.3(3)   & 49(7)    &  1.2 &  S &  0.46$\pm$0.17  &     & E\,3674      		      & 0.32 \\
$^{*}$38 & 17.76   & 8 52 22.09(8)  &   11 39 12(1)     & 49(13)   &  1.7 &  T &  0.76$\pm$0.22  &     & 	      &  \\
 39 &  3.66   &   8 51 08.95(2)     &   11 50 43.5(2)   & 48(7)    &  1.1 &  T &  0.96$\pm$0.09  &     & faint (extended)  	              & 0.4 \\
 40 &  5.76   &   8 51 36.20(2)     &   11 44 46.0(3)   & 47(7)    &  1.2 &  T &  0.74$\pm$0.13  &     & 	      		      &      \\
 41 &  2.01   &   8 51 28.90(1)     &   11 48 07.4(2)   & 47(7)    &  1.2 &  T &  0.80$\pm$0.13  &     &      	      		      &      \\
 42 &  5.41   &   8 51 30.30(2)     &   11 54 39.7(3)   & 47(7)    &  1.1 &  T &  0.63$\pm$0.15  &     & 		      		      &      \\
 43 &  8.09   &   8 51 55.81(2)     &   11 50 44.5(4)   & 47(7)    &  1.2 &  T &  0.65$\pm$0.15  &     & faint (extended)    	              & 0.7 \\
 44 & 16.16   &   8 52 05.80(6)     &   11 37 13(1)     & 46(11)   &  1.5 &  T &  0.31$\pm$0.26  & 58  &       		      &      \\
 45 &  6.26   &   8 50 58.97(2)     &   11 51 37.4(3)   & 45(7)    &  1.3 &  T &  0.67$\pm$0.11  & 39  &  E\,1268$^h$      & 1.86 \\
 46 &  7.30   &   8 50 53.29(2)     &   11 49 43.5(4)   & 42(7)    &  1.1 &  T &  0.55$\pm$0.16  &     & faint$^e$ & 1.2     \\
 47 &  7.48   &   8 51 50.28(2)     &   11 46 07.6(4)   & 42(7)    &  1.1 &  S &  0.10$\pm$0.17  & 52  & S\,1237$^d$   	      & 0.91 \\
 48 &  2.76   &   8 51 19.72(1)     &   11 52 10.9(2)   & 42(6)    &  0.94 &  S & $-$0.40$\pm$0.19  & 38  & S\,1070 	      & 0.05 \\
 49 &  4.36   &   8 51 36.06(2)     &   11 46 33.6(2)   & 41(6)    &  1.1 &  S & $-$0.57$\pm$0.18  & 50  & S\,1242   & 0.45 \\
 50 & 19.69   &   8 52 27.01(8)     &   11 37 35(1)     & 39(11)   &  1.4 &  S &  0.76$\pm$0.24  &     & E\,1939$^{g}$  & 4.57  \\
  \hline										  
\end{tabular}									  
\end{table*}	

\begin{table*}	
\begin{tabular}{rrllrcrrlll}
\hline \hline
CX  & $d_{\rm aim}$ & \multicolumn{1}{c}{$^{**}\alpha_{\rm J2000}$}  & \multicolumn{1}{c}{$^{**}\delta_{\rm J2000}$} & counts & ctr & $B$ & \multicolumn{1}{c}{$HR$} & X   & opt  & $d_{\rm XO}$  \\
    & \multicolumn{1}{c}{(\arcmin)}  &  \multicolumn{1}{c}{($^h$,$^m$,$^s$)}  &  \multicolumn{1}{c}{($^{\circ}$,\arcmin,\arcsec)}   &    & (c\,ks$^{-1}$) &  &  &      &      & (\arcsec) \\ \hline
 51 & 11.21   &  8 52 08.47(4)       &   11 47 59.2(8)  &  38(9)      &   0.99	     & T  &  0.86$\pm$0.15    &     &                     &      \\
 52 &  2.92   &  8 51 11.73(2)       &   11 50 25.8(2)  &  38(6)      &   0.89	    & T  &  0.90$\pm$0.13    &     & 	       		&      \\
 53 &  7.73   &  8 50 56.50(3)       &   11 45 23.3(4)  &  36(7)      &   1.0	    & T  &  0.24$\pm$0.20    & 24  & E\,362$^{h}$        & 2.27 \\
$^b$54 & 15.51   & 8 52 16.86(8)     &   11 41 21(1)    &  35(9)      &   1.1 	    & S  & $-$0.03$\pm$0.24    & 56$^{c}$ & E\,207$^{f}$ & 2.05 \\
 55 &  6.62   &  8 51 33.80(2)       &   11 43 27.8(4)  &  32(6)      & 0.83 	    & S  &  0.35$\pm$0.21    &     & E\,2684    & 0.88 \\
 56 &  8.47   &  8 51 04.60(4)       &   11 42 23.7(5)  &  31(6)      & 0.81 	    & T  &  0.48$\pm$0.24    &     & E\,2050    & 0.47 \\
 57 &  2.78   &  8 51 27.21(1)       &   11 46 57.3(2)  &  30(5)      & 0.75 	    & T  &  0.41$\pm$0.22    & 16  & E\,2164    & 0.05 \\
 58 & 10.02   &  8 50 55.00(5)       &   11 56 50.1(6)  &  29(6)      & 0.91 	    & S  & $-$0.21$\pm$0.27    &   & E\,763     & 0.44 \\
 59 &  7.08   &  8 50 57.55(4)       &   11 52 51.5(4)  &  28(6)      & 0.74 	    & T  &  0.65$\pm$0.19    & 36  &            &      \\
 60 &  6.45   &  8 51 44.50(3)       &   11 45 47.0(4)  &  28(6)      & 0.78 	    & T  &  \multicolumn{1}{c}{1.00}    &     & &      \\
 61 &  6.73   &  8 51 26.77(3)       &   11 56 13.1(4)  &  27(6)      & 0.70 	    & S  &  0.27$\pm$0.25   &     & E\,742      & 0.83 \\
 62 &  6.30   &  8 51 06.31(2)       &   11 54 19.4(3)  &  27(5)      & 0.69 	    & S  &  0.17$\pm$0.25   &     & E\,683$^h$  & 1.47 \\
 63 &  8.26   &  8 51 25.36(4)       &   11 57 47.5(4)  &  26(6)      & 0.68 	    & T  &  0.62$\pm$0.23   &     & faint       & 0.6 \\
 64 &  4.41   &  8 51 18.81(2)       &   11 53 49.9(2)  &  25(5)      & 0.59 	    & T  &  0.62$\pm$0.23   &     & E\,2306  & 0.09 \\
 65 &  4.43   &  8 51 20.86(2)       &   11 53 56.4(3)  &  24(5)      & 0.56 	    & T  &  0.84$\pm$0.21   &     & E\,2307  & 0.19 \\
 66 &  7.89   &  8 51 35.22(3)       &   11 56 51.7(7)  &  24(6)      & 0.60 	    & T  &  0.51$\pm$0.27   &     &         	      &      \\
 67 & 10.27   &  8 51 35.59(5)       &   11 39 44.8(5)  &  24(6)      & 0.70 	    & S  &  0.13$\pm$0.26   &     & S\,1203$^h$      & 2.44 \\
 68 &  5.28   &  8 51 01.55(2)       &   11 49 27.5(2)  &  24(5)      & 0.62 	    & T  & $-$0.79$\pm$0.25   & 44  & S\,2214   & 0.80 \\
 69 &  3.40   &  8 51 09.39(2)       &   11 49 02.1(2)  &  24(5)      & 0.54 	    & T  &  0.28$\pm$0.26   &     & faint   	      & 0.8 \\
 70 &  5.17   &  8 51 44.05(2)       &   11 48 49.7(3)  &  23(5)      & 0.56 	    & S  &  0.61$\pm$0.24   &     & E\,2208 	       		      & 0.81 \\
 71 &  6.02   &  8 51 38.45(2)       &   11 54 15.2(3)  &  22(5)      & 0.53 	    & S  &  0.47$\pm$0.23   &     & E\,2807  	       		      & 0.88 \\
 72 &  5.16   &  8 51 02.12(4)       &   11 49 01.3(6)  &  20(5)      & 0.50 	    & S  & $-$0.43$\pm$0.33   &  &   S\,773       	      & 0.63 \\
 73 & 10.73   &  8 51 43.13(6)       &   11 40 00.1(8)  &  19(5)      & 0.57 	    & T  &  0.77$\pm$0.27   &     & E\,2650$^h$	   	      & 3.71 \\
 74 &  2.00   &  8 51 15.85(3)       &   11 48 37.2(5)  &  19(4)      & 0.42 	    & T  &  0.15$\pm$0.32   &     & faint       & 0.2 \\
 75 &  6.48   &  8 50 56.89(4)       &   11 50 26.4(6)  &  19(5)      & 0.46 	    & T  &  0.60$\pm$0.25   &     & 	       		      &      \\
 76 &  1.84   &  8 51 17.08(3)       &   11 48 27.1(5)  &  18(4)      & 0.40 	    & S  &  0.05$\pm$0.34   &     & E\,1720  & 0.13 \\
 77 &  4.25   &  8 51 32.67(4)       &   11 45 59.9(5)  &  18(4)      & 0.42 	    & S  &  0.06$\pm$0.34   &     & E\,394   	       		      & 0.30 \\
 78 &  1.89   &  8 51 19.21(3)       &   11 47 54.9(5)  &  16(4)      & 0.74 	    & S  & $-$0.19$\pm$0.37   &     & S\,1009       & 0.08 \\
 79 &  4.78   &  8 51 17.09(4)       &   11 54 05.7(6)  &  15(4)      & 0.38 	    & T  &  0.60$\pm$0.35   &     & E\,2310        		      & 0.25 \\
 80 & 12.16   &  8 52 12.80(6)       &   11 49 49.6(9)  &  15(6)      & 0.36 	    & S  &  0.11$\pm$0.40   &     & E\,2228$^{g}$  		      & 2.74 \\
 81 &  3.01   &  8 51 28.37(3)       &   11 46 49.8(5)  &  15(4)      & 0.34 	    & T  & $-$0.38$\pm$0.40   &     & S\,996       	       	      & 0.10 \\
 82 &  3.71   &  8 51 11.42(4)       &   11 47 11.2(5)  &  14(4)      & 0.34 	    & T  & $-$0.09$\pm$0.37   &     & E\,1208      	       		      & 0.45 \\
 83 &  4.54   &  8 51 05.41(4)       &   11 48 10.7(5)  &  14(4)      & 0.34 	    & H  &  \multicolumn{1}{c}{1.00} & &   		      &      \\
 84 &  4.52   &  8 51 40.80(3)       &   11 48 13.5(5)  &  14(4)      & 0.35 	    & T  &  0.88$\pm$0.29   &     & 		       		      &      \\
 85 &  6.68   &  8 51 09.77(5)       &   11 43 43.5(8)  &  14(4)      & 0.35 	    & T  &  0.79$\pm$0.34   &     & 		       		      &      \\
 86 &  7.15   &  8 50 58.72(5)       &   11 45 36.7(7)  &  14(4)      & 0.36 	    & T  &  0.92$\pm$0.31   &     & E\,2135$^h$ 	       		      & 2.74      \\
 87 &  8.56   &  8 51 08.73(6)       &   11 57 21.0(8)  &  14(4)      & 0.35 	    & T  &  0.86$\pm$0.31   &     &   		      &      \\
 88 &  1.14   &  8 51 19.05(4)       &   11 50 05.7(5)  &  14(4)      & 0.33 	    & S  & $-$0.55$\pm$0.39   & 41  & S\,1045     	      & 0.27 \\
 89 &  2.63   &  8 51 21.67(4)       &   11 52 09.0(5)  &  14(4)      & 0.34 	    & T  &  0.43$\pm$0.37   & 38  & E\,2270        		      & 0.52 \\
 90 &  6.98   &  8 51 19.94(5)       &   11 56 29.3(7)  &  13(4)      & 0.34 	    & T  &  0.92$\pm$0.33   &     &  	                      &      \\
 91 &  5.77   &  8 51 41.60(4)       &   11 53 08.2(6)  &  13(4)      & 0.32 	    & T  &  0.61$\pm$0.35   &     &  	                      &      \\
 92 &  1.61   &  8 51 25.04(3)       &   11 48 00.8(5)  &  13(4)      & 0.31 	    & T  &  0.85$\pm$0.33   &     &  		                      &      \\
$^{*}$93 & 9.50   & 8 51 02.73(6)   &    11 57 38.4(8)    &  13(4)      & 0.33 	    & T   &  0.71$\pm$0.38   &    &  	                      &      \\
 94 &  3.05   &  8 51 34.27(3)       &   11 50 55.1(5)  &  13(4)      & 0.29 	    & T  &  0.26$\pm$0.35   &     & S\,1281         	      & 0.48 \\
 95 &  2.65   &  8 51 19.94(3)       &   11 47 00.8(5)  &  12(3)      & 0.27 	    & S  &   &     & S\,997      	      & 0.35 \\
 96 &  3.07   &  8 51 26.98(3)       &   11 46 38.0(5)  &  12(3)      & 0.27 	    & T  &   &     &  	       		      &      \\
 97 &  4.96   &  8 51 35.79(4)       &   11 53 24.9(5)  &  11(3)      & 0.28 	    & T  & 0.67$\pm$0.39   &     &        		      &      \\ 
 98 &  4.94   &  8 51 11.06(4)       &   11 45 35.3(5)  &  11(3)      & 0.26 	    & T  &                   &     & faint (extended)    	      & 0.8 \\
 99 &  7.57   &  8 50 52.25(5)       &   11 49 03.0(6)  &  11(4)      & 0.30 	    & T  & 0.70$\pm$0.40      &     & 	       		      &      \\
100 &  8.36   &  8 51 50.37(6)       &   11 44 31.1(9)  &  11(4)      & 0.64 	    & T  &    &     & E\,3668  	       		      & 1.04 \\
101 &  1.63   &  8 51 18.72(3)       &   11 50 46.4(5)  &  11(3)      & 0.25 	    & T  &    &     &	       		      &      \\
102 &  3.56   &  8 51 28.42(3)       &   11 52 52.1(5)  &  11(3)      & 0.25 	    & T  &    &     & faint (extended) & 2.9     \\
103 &  4.89   &  8 51 14.27(4)       &   11 53 55.1(5)  &  10(3)      & 0.25 	    & T  & 0.83$\pm$0.40   &     & 	       		      &      \\
104 &  1.42   &  8 51 18.29(3)       &   11 50 19.7(5)  &  10(3)      & 0.23 	    & S  &  & 41  & S\,1050  	              & 0.21 \\
$^{*}$105 & 8.77   & 8 51 46.20(4)  &    11 42 50.2(8)  &  10(4)      & 0.26 	    & T   &    &     &      		      &      \\
106 &  6.03   &  8 51 07.09(4)       &   11 54 07.6(6)  &  10(3)      & 0.24 	    & T  &    &     & 	      		      &      \\
107 &  5.34   &  8 51 29.91(4)       &   11 44 28.3(6)  &   9(3)      & 0.25 	    & T  & 0.83$\pm$0.40    &     &       		      &      \\
108 &  5.67   &  8 51 25.99(4)       &   11 43 55.4(6)  &   9(3)      & 0.23 	    & T  &    &     &       		      &      \\
109 &  8.29   &  8 51 56.10(5)       &   11 51 27.5(7)       &   9(3)      & 0.24 	    & S  &    &     & S\,1466     & 1.36 \\
\hline								     		  
\end{tabular}							     		  
\end{table*}							     		  
	     		  
\begin{table*}	
\begin{tabular}{rrllrcrrlll}
\hline \hline
CX  & $d_{\rm aim}$ & \multicolumn{1}{c}{$^{**}\alpha_{\rm J2000}$}  & \multicolumn{1}{c}{$^{**}\delta_{\rm J2000}$} & counts & ctr & $B$ & \multicolumn{1}{c}{$HR$} & X   & opt  & $d_{\rm XO}$  \\
    & \multicolumn{1}{c}{(\arcmin)}  &  \multicolumn{1}{c}{($^h$,$^m$,$^s$)}  &  \multicolumn{1}{c}{($^{\circ}$,\arcmin,\arcsec)}   &    & (c\,ks$^{-1}$) &  &  &      &      & (\arcsec) \\ \hline
110 &  2.30   & 8 51 23.23(4)         &   11 47 14.8(5)   &   9(3)        & 0.27  & T  &    &     &       		      &      \\
111 &  0.72   & 8 51 22.95(3)         &   11 48 50.0(5)   &   8(3)        & 0.18  & S  &    & 46  & S\,1024  	      & 0.32 \\
112 &  0.61   & 8 51 23.12(3)         &   11 48 56.2(5)   &   8(3)        & 0.18  & H  &    & 46  &  	      		      &      \\
113 &  2.43   & 8 51 16.58(4)         &   11 51 22.4(5)   &   8(3)        & 0.17  & T  &    &     & 		      		      &      \\
114 &  3.75   & 8 51 20.79(4)         &   11 45 50.4(5)   &   8(3)        & 0.21  & T  &    &     & 		      		      &      \\
115 &  4.93   & 8 51 08.16(4)         &   11 52 50.8(5)   &   8(3)        & 0.18  & T   &    &     & 		      		      &      \\
$^{*}$116 & 7.26 & 8 51 52.03(5)  &       11 51 10.6(6)    &   7(3)        & 0.19  & T   &    &     & 	      		      &      \\
117 &  2.30   & 8 51 31.59(4)        &    11 50 32.6(5)   &   7(3)        & 0.16  & S  &    &     & E\,542    		      & 0.13 \\
118 &  3.98   & 8 51 07.85(4)        &    11 48 09.8(5)   &   7(3)        & 0.16  & T  &    &     & S\,1013   		      & 0.27 \\
119 &  3.79   & 8 51 28.45(4)        &    11 53 06.6(5)   &   7(3)        & 0.16  & T  &    &     & 	                      &      \\
120 &  4.37   & 8 51 31.83(4)        &    11 45 44.4(6)   &   7(3)        & 0.18  & T   &    & 53  & faint                  & 0.6 \\ 
121 &  6.96   & 8 51 29.56(6)        &    11 56 19.5(7)   &   7(3)        & 0.32  & S  &    &      &  	                      &      \\
$^{*}$122 &  8.43   & 8 50 49.04(5) &     11 48 16.5(7)   &   7(3)        & 0.18  & T   &    &     &  	                      &      \\
$^{*}$123 &  5.18   & 8 51 41.38(4) &     11 46 56.1(7)   &   6(3)        & 0.17  & T   &    &     &  	                      &      \\
$^{*}$124 &  8.61   & 8 51 53.94(4) &     11 53 42.2(6)   &   6(3)        & 0.16  & T   &    &     & faint              & 2.3 \\
$^{*}$125 &  6.34   & 8 51 48.27(5) &     11 51 02.8(7)   &   6(3)        & 0.14  & T  &    &     &  	                      &      \\
126 &  4.23   & 8 51 25.92(4)        &    11 53 43.3(5)   &   6(2)        & 0.24  & T  &    &     &  	                      &      \\
127 &  2.61   & 8 51 29.29(4)        &    11 47 25.0(5)   &   6(2)        & 0.13  & S  &    &     & E\,3701  	                      & 0.49 \\
128 &  5.40   & 8 51 01.15(3)        &    11 50 02.4(7)   &   5(2)        & 0.13  & T  &    &     & 	                      &      \\
129 &  5.13   & 8 51 43.09(4)        &    11 51 05.4(7)   &   5(2)        & 0.14  & T  &    &     & E\,556$^h$   & 3.27 \\
$^{*}$130 &  7.67   & 8 50 51.76(4) &     11 49 21.1(7)   &  5(2)         & 0.14  & T   &    &     &                       &      \\
$^{*}$131 &  1.14   & 8 51 24.30(4) &     11 50 39.1(5)   &  5(2)         & 0.11  & T   &    &     &                       &      \\
132 &  4.13         & 8 51 11.57(4) &     11 46 32.5(5)   &  5(2)         & 0.11  & T   &    &     &                       &      \\
$^{*}$133 &  3.57   & 8 51 37.38(4) &     11 50 17.0(5)   &  5(2)         & 0.11  & T   &    &     &                       &      \\
134 &  5.06         & 8 51 09.09(4)   &   11 53 16.1(5)   &  4(2)         & 0.10  & T  &    &     &                        &      \\
135 &  6.86         & 8 51 35.73(4)   &   11 43 25.7(5)   &  4(2)         & 0.11  & T   &    &     &                      &      \\
$^{*}$136 &  6.77   & 8 50 55.66(6) &     11 48 40.5(7)   &  4(2)         & 0.10  & H  &    &     & faint &  0.2    \\
$^{*}$137 &  3.77   & 8 51 15.83(4) &     11 46 13.7(5)   &  4(2)         & 0.26  & T   &    &     &                   &      \\
$^{*}$138 &  2.93   & 8 51 11.24(4) &     11 49 09.1(5)   &  4(2)         & 0.09  & T   &    &     &                       &      \\
$^{*}$139 &  6.69   & 8 51 50.33(5) &     11 48 55.9(7)   &  4(2)         & 0.09  & T   &    &     & F\,4345$^d$            	              & 0.61 \\
140 &  5.52   & 8 51 03.11(4)       &     11 46 59.5(5)   &  4(2)         & 0.08  & H  &    &     & 		      &      \\
141 &  3.11   & 8 51 35.19(3)       &     11 50 31.8(5)   &  3(2)         & 0.07  & S  &    &     & F\,3925$^d$ 	      & 0.59 \\
$^{*}$142 &  1.78   & 8 51 15.89(3)&      11 49 23.3(5)   &  3(2)         & 0.06  & T   &    &     &               		      &      \\
143 &  1.93         & 8 51 24.93(3)     & 11 47 40.5(5)   &  3(2)         & 0.07  & T   &    &     &               		      &      \\
144 &  2.68         & 8 51 23.17(4)     & 11 46 52.3(5)   &  3(2)         & 0.07  & T   &    &     &                 		      &      \\
$^{*}$145 &  3.14   & 8 51 12.88(4)&      11 47 39.5(5)   &  3(2)         & 0.07  & T   &    &     & faint  & 1.0 \\
146       & 15.94   & 8 52 28.02(6)&      11 50 55(1)     &  40(10)$^{a}$ & 0.87  & S   &  0.64$\pm$0.19  &     &        &      \\  
$^{*}$147 & 13.88   & 8 52 18.22(7)&      11 46 15(1)     &  27(8)$^{a}$  & 0.54  & S   &    &     & faint & 3.8     \\
$^{*}$148       & 15.99   & 8 52 20.97(9)&      11 42 07(1)     &  25(7)$^{a}$  & 0.87 & H   & \multicolumn{1}{c}{1.00}   &     &    	      &      \\
$^{*}$149 &  5.99   & 8 51 38.52(4)&      11 54 12.3(6)   &  3(2)$^{a}$   & 0.06 & H   &    &  & E\,2807$^h$            & 2.33    \\ 
$^{*}$150 & 11.75   & 8 52 01.09(9)&      11 42 22(1)     &  10(4)$^{a}$  & 0.26 & S   &  0.59$\pm$0.28  &     &              &      \\ 
$^{*}$151 &  7.70   & 8 51 49.82(4)&      11 45 28.9(8)   &  5(2)$^{a}$   & 0.11 & S   &    &     &   	      &      \\   
$^{*}$152 &  6.39   & 8 50 57.46(4)&      11 50 45.1(7)   &  4(2)$^{a}$   & 0.11 & S   &    &     & F\,2776$^h$ & 1.97 \\
$^{*}$153 &  4.79   & 8 51 23.51(4)&      11 44 45.7(5)   &  4(2)$^{a}$   & 0.10 & S   &    &     & E\,1178  & 0.74 \\  
154       &  4.64   & 8 51 04.24(4)&      11 49 07.4(5)   &  4(2)$^{a}$   & 0.10 & S   &    &     &	       &   \\ 
155       &  4.74   & 8 51 42.45(4)&      11 49 54.3(6)   &  4(2)$^{a}$   & 0.09 & S   &    &     & S\,1272$^h$ 	& 2.05 \\
$^{*}$156 &  2.49   & 8 51 32.13(3)&      11 50 42.3(5)   &  3(2)$^{a}$   & 0.07 & S   &    &     &      &      \\ 
157       &  3.85   & 8 51 18.01(3)&      11 45 54.1(5)   &  2(1)$^{a}$   & 0.04 & S   &    &     & S\,986 & 0.35 \\
 $^{*}$158       &  3.16   & 8 51 35.37(3)&      11 48 32.8(5)   &  2(1)$^{a}$   & 0.04 & S   &    &     & E\,3534   & 1.03 \\
\hline       					 
\end{tabular}					    
				    
$^*$ Source is detected with the significance threshold set to
$10^{-6}$ but not with the threshold set to $10^{-7}$\\ $^{**}$ All
positions are corrected for a systematic offset with respect to the
EIS optical positions which is $\Delta \alpha = \alpha_{\rm X} -
\alpha_{\rm opt} = -$0\farcs23 and $\Delta \delta = \delta_{\rm X} -
\delta_{\rm opt} = -$0\farcs20. The errors on the positions that we
give here are the 1-$\sigma$ {\em wavdetect} errors including the
corrections for small counts numbers. They are often much smaller than
the uncertainty in the absolute positions, i.e.\, than the uncertainty
in the offset values ($\sim$0\farcs3).\\
				    
\end{table*}					
				    
\begin{table*}					  
\begin{tabular}{cc}				       
 & \mbox{~~~~~~~~~~~~~~~~~~~~~~~~~~~~~~~~~~~~~~~~~~~~~~~~~~~~~~~~~~~~~~~~~~~~~~~~}\\					
\end{tabular}					 
				    
Notes on individual sources:\\ 					   
$^{a}$ For sources CX\,148 and CX\,149, and for CX\,146, CX\,147 and					
CX\,150--158, that were only detected in the hard and soft band					   
respectively, the count(rate)s refer to the hard and soft band.\\				       
$^{b}$ CX\,54: properties may be affected by bad pixel region.\\				    
$^{c}$ CX\,4, CX\,13, CX54: identification with ROSAT source is uncertain.\\				     
$^{d}$ CX\,6, CX\,24, CX\,47, CX\,139, CX\,141: the optical				       
       counterparts are not included in the EIS catalogue, or (for				      
       S\,1237) the EIS position is clearly offset from the centre of					 
       the optical source. For these sources, $d_{\rm OX}$ is the				     
       distance between the optical position from Fan et al., and the X-ray				       
       position (corrected for the systematic offset with respect to				       
       the Fan et al.\,catalogue: $\Delta \alpha =				      
       $0\farcs11$\pm$0\farcs35, $\Delta \delta =				     
       $0\farcs23$\pm$0\farcs29).\\				      
$^e$ CX\,46: besides the faint counterpart, two optical sources lie				       
     inside the source region but further away from the X-ray 				     
     source:					
     E\,1243 at 4\farcs2, and					 
     S\,778 at 3\farcs7.\\				     
$^f$ CX\,30, CX\,34, CX\,35, CX\,54: ACIS-S sources; optical					
     counterpart only found when we use 2-$\sigma$ {\em wavdetect}				      
     errors.  For CX\,30 and CX\,35 the counterpart is only found for					 
     the total-band source. For CX\,34 and CX\,54 the counterpart is only found for the soft source; therefore, $d_{\rm XO}$ refers to the distance between the optical source and the soft source. \\					   
$^g$ CX\,26, CX\,50, CX\,80: ACIS-S sources; optical counterpart only					 
     found when we use 3-$\sigma$ {\em wavdetect} errors. For CX\,50					
     and CX\,80 the counterpart is only found for the soft source, therefore, $d_{\rm XO}$ refers to the distance between the optical source and the soft source.\\					
$^h$ CX\,32, CX\,45, CX\,53, CX\,62, CX\,67, CX\,73, CX\,86, CX\,129,					 
     CX\,149, CX\,152, CX\,155: optical counterpart lies outside errorbox but					 
     inside the source region (we only indicate cases for				    
     which $d_{\rm OX} < 4$\arcsec). \\					   
				    
\end{table*}					

\setcounter{table}{3}					 

\begin{table*} 					  
\caption{Positions of faint optical counterparts as estimated from the
EIS image.  We give the Chandra identification (CX) and, if
appropriate, the corresponding ROSAT identification (X). For ROSAT
sources, we indicate in the last column whether the present
counterpart is new (N). The '?' marks an identification that is
uncertain because of a relatively large offset between the Chandra and
ROSAT position; see the discussion in Sect.~\ref{rosatid}.
\label{fc}}				       
\begin{tabular}{rllll}					  
\hline 					   
\hline 					   
CX & \multicolumn{1}{c}{X} &  $\alpha_{\rm J2000}$   & $\delta_{\rm J2000}$       & \\	
   & & ($^{h}$,$^{m}$,$^{s}$) & ($^{\circ}$,\arcmin,\arcsec) & \\			    
\hline 
  4 & 15? & 8 50 57.10 & 11 46 07.7 & N \\ 				       
  8 & 32 & 8 51 27.12 & 11 55 24.5  & N  \\				      
 39 &    & 8 51 08.94 & 11 50 43.8  &   \\				      
 43 &    & 8 51 55.77 & 11 50 44.8  &   \\ 				       
 46 &    & 8 50 53.37 & 11 49 43.9  &   \\ 				       
 63 &    & 8 51 25.39 & 11 57 47.1  &   \\ 				       
 69 &    & 8 51 09.43 & 11 49 01.6  &   \\ 				       
 74 &    & 8 51 15.86 & 11 48 37.2  &   \\ 				       
 98 &    & 8 51 11.09 & 11 45 35.9  &   \\ 				       
102 &    & 8 51 28.60 & 11 52 53.3  &   \\ 				       
120 & 53 & 8 51 31.82 & 11 45 43.8  &   \\ 					 
124 &    & 8 51 53.80 & 11 53 41.2  &   \\ 				       
136 &    & 8 50 55.67 & 11 48 40.5  &   \\ 				       
145 &    & 8 51 12.90 & 11 47 40.4  &   \\ 				       
147 &    & 8 52 18.10 & 11 46 18.7  &   \\ 				       
\hline					  
\end{tabular}					 
\end{table*}

\end{document}